\documentclass[aps,prd,groupedaddress,nofootinbib,letterpaper,superscriptaddress]{revtex4}

\usepackage{amsmath}
\usepackage{amsfonts}
\usepackage{amssymb}

\usepackage{braket,mathtools}
\usepackage{mathrsfs}
\usepackage{stmaryrd}

\usepackage{hyperref}
\usepackage{color}

\DeclarePairedDelimiter\abs{\lvert}{\rvert}%

\newcommand{\call}{{\cal L}}
\newcommand{\calo}{{\cal O}}
\newcommand{\scri}{{\mathscr I}}
\newcommand{\beq}{\begin{equation}}
\newcommand{\eeq}{\end{equation}}
\newcommand{\hf}{\frac{1}{2}}
\newcommand{\xpp}{x^{\prime\prime}}
\newcommand{\vxpp}{\vec x^{\prime\prime}}

\begin{document}
\begin{titlepage}

\hfill NSF-KITP-15-133

\bigskip\bigskip

\title{Diffeomorphism-invariant observables and their nonlocal algebra} 
\preprint{NSF-KITP-15-133}

\author{William Donnelly}
\email{donnelly@physics.ucsb.edu}
\affiliation{Department of Physics, University of California, Santa Barbara, CA 93106}

\author{Steven B. Giddings}
\email{giddings@physics.ucsb.edu}
\affiliation{Department of Physics, University of California, Santa Barbara, CA 93106}
\affiliation{Kavli Institute of Theoretical Physics, University of California, Santa Barbara, CA 93106}

\begin{abstract}
Gauge-invariant observables for quantum gravity are described, with explicit constructions given primarily to leading order in Newton's constant, analogous to and extending constructions first given by Dirac in quantum electrodynamics.  These can be thought of as operators that create a particle, together with its inseparable gravitational field, and reduce to usual field operators of quantum field theory in the weak-gravity limit; they include both Wilson-line operators, and those creating a Coulombic field configuration.  We also describe operators creating the field of a particle in motion; as in the electromagnetic case, these are expected to help address infrared problems.  An important characteristic of the quantum theory of gravity is the algebra of its observables.  We show that the commutators of the simple observables of this paper are nonlocal, with nonlocality becoming significant in strong field regions, as predicted previously on general grounds.  
\end{abstract}

\maketitle

\end{titlepage}

\newpage
\thispagestyle{empty}
\tableofcontents
\newpage

\section{Introduction}

Gravity is widely believed to be described by a quantum-mechanical theory, rather than one that requires an extension or modification of quantum mechanics.  If this is the case, the structure of quantum mechanics (suitably generalized -- see \cite{UQM}) imparts certain rigid features to the theory of quantum gravity.  One basic aspect is the existence of quantum observables, which are gauge-invariant, Hermitian operators acting on the Hilbert space of states of the theory.  An outstanding problem is to understand properties of these observables.

While a complete discussion of observables in quantum gravity must obviously await more complete understanding of its Hilbert space and dynamics, we already have a good deal of information if any such more complete formulation is to match onto a quantum version of Einstein's theory in weak field regimes.  One can study the properties of quantum observables in these regimes.  Moreover, such properties are likely to provide further information about the mathematical structure of the theory in the strong-gravity regime.   One of the challenges of quantum gravity is specifically to formulate gauge-invariant observables that reduce to the usual observables of local quantum field theory (LQFT) in the weak-field limit.  

In order to define gauge-invariant observables that reduce to those of LQFT, ordinarily a relational approach is taken\cite{Dewi1,Dewi2,TsWo,Rove,Maro,Ditt,PoSa,Giddings2005,DiTa,GaGi,GiSl,Khavkine:2011kj,Bonga:2013uha}, where for example a particle or field operator is localized with respect to some features of the state, or with respect to another particle or field operator.  Some examples of such constructions are given in \cite{Giddings2005}.  We also expect that there should be observables that act on a state of the system, say the vacuum, and create or annihilate a particle, as in LQFT.  These are expected to be the simplest operators reducing to simple operators of a non-gravitational LQFT.  A key point, however, is that such observables must {\it also} create the gravitational field of the particle, in order to be gauge invariant (that is, satisfy the constraints).  Such operators have been constructed in gauge theories -- going back to the work of Dirac\cite{Dirac1955} -- but not, to our knowledge, in gravity.

Indeed, one way to think of constructing such operators has a close parallel to construction of other relational operators\cite{Giddings2005}; we can demand that the point at which a field operator acts is a fixed geodesic distance from a fixed feature (or ``platform"), which we may take to approach infinity.  Such a specification will be diffeomorphism invariant for diffeomorphisms vanishing at infinity.  An example, in anti de Sitter (AdS) space, is to base such coordinates on spatial infinity, which there serves as the ``platform."  An equivalent way to describe these constructions defines operators by working in a specific gauge, {\it e.g.} using Gaussian normal coordinates based on the asymptotic platform.  Such a construction was considered for AdS in \cite{Heemskerk2012,Kabat:2012hp,Kabat2013}; a related construction appears in \cite{Bodendorfer:2015aca}, based on earlier work \cite{Duch:2014hfa}.

These observables, acting on the vacuum, create both the quantum associated with the field operator, and a non-trivial gravitational field.  The field for the ``Wilson-line" observables we have just described is a singular gravitational string.  Such a string, once created, is expected to decay to a more natural, less singular gravitational configuration.  An approach to deriving the operator that directly creates such a configuration is to {\it average} over the directions of the gravitational string.  Working in the linear theory, we will find that such a procedure indeed produces an operator that creates the gravitational analog of the Coulomb field, namely the linearized Schwarzschild solution; a parallel construction works for quantum electrodynamics (QED), producing the Dirac dressing.  These operators, and other operators that we construct taking into account possible motion of the particles associated with the field operator, give simple example of diffeomorphism-invariant observables in quantum gravity, which we explicitly construct to leading order in an expansion in Newton's constant $G$.  

Another key question regarding gravitational observables is the algebra that they obey.  In LQFT, the field algebra closely mirrors the underlying manifold structure\cite{Haag} and provides a precise characterization of locality, through commutativity of observables associated with spacelike-separated regions.  An important question for quantum gravity, which appears critical to inferring its underlying structure and its interplay with locality, is that of determining the structure of the algebra obeyed by its observables\cite{SGalg}.  Since we are able to construct such gauge invariant observables to leading order in $G$, we are able to infer the leading-order structure of this algebra, and we do so by explicitly calculating commutators of the operators we have just described.  We find that these operators do not have local commutation relations,\footnote{Here, we differ from claims of \cite{Kabat2013}, but for reasons that can and will be explained in the analogous and simpler case of QED.} due to the gravitational dressings that we have described, and specifically they have significant departures  from the commutators of LQFT in regions previously characterized by the {\it locality bounds} of 
\cite{GiLia,GiLib,LQGST}.  These basic algebraic properties of the theory appear to be the universal weak-field behavior of any quantum gravity theory that matches Einstein's in the weak field limit, and thus should furnish important information about the more complete theory of quantum gravity.

In outline, the next section reviews and extends the discussion of gauge-invariant observables in QED, to set the stage for gravity.  Section III describes construction of both ``Wilson-line" and ``Coulomb" diffeomorphism-invariant observables in gravity, as well as a generalization valid for moving particles, and performs simple checks that these operators create the physically-expected gravitational fields.  Section IV turns to the question of the algebraic structure, first in  QED, where we derive nonzero commutators for various dressings and explicitly reconcile a conflict with \cite{Kabat2012}, by also working out commutators using Dirac brackets.  We then exhibit the non-trivial and nonlocal commutators of various of the 
gravitational field dressings.  Section V provides conclusions and discussion of further directions.  There are also three appendices: two with basic formulas for quantization of QED and of linearized gravity, and one with other useful formulas.

\section{Gauge-invariant observables for QED}

We begin by reviewing and extending discussions of gauge-invariant observables in QED, which we will take to be coupled to a scalar $\phi$ with charge $q$,
\beq
\call_{\rm QED} = -\frac14 F^{\mu \nu} F_{\mu \nu} - \frac{1}{2\alpha} (\partial_\mu A^\mu)^2-\abs{D_\mu \phi}^2 - m^2 \abs{\phi}^2\ ,
\eeq
where $D_\mu = \partial_\mu - iq A_\mu$.
Gauge transformations act as
\begin{eqnarray}
A_\mu(x) & \rightarrow & A_\mu(x) - \partial_\mu\Lambda(x) \label{gaugeXM}\\
\phi(x) & \rightarrow & e^{-iq\Lambda(x)}\phi(x), \label{Fieldxm}
\end{eqnarray}
where the gauge transformation parameter $\Lambda(x)$ vanishes at infinity.  The parameter $\alpha$ is a gauge-fixing (more precisely, breaking) parameter; special choices are $\alpha=0$, Lorenz or Landau gauge, $\alpha=1$, ``Feynman gauge," and $\alpha=\infty$, restoring gauge symmetry.  Further conventions and useful formulas for QED appear in Appendix \ref{app:qed}.  

The na\"\i ve expectation that $\phi$ acts on the vacuum to create particles is confounded by the fact that $\phi$ is not gauge invariant.  However, one may ``dress" $\phi$ to make a gauge invariant, if one defines
\beq
\Phi(x) = V(x) \phi(x)
\eeq
where the dressing $V$ transforms as $V(x) \to e^{i q \Lambda(x)} V(x)$.
Following 
 \cite{Dirac1955,Bucholz1982,Steinmann1983,Steinmann2004}  such $V$'s can be found in the form 
\begin{equation} \label{V}
V(x) = \exp\left( i q \int d^4x' f^\mu(x,x') A_\mu(x') \right).
\end{equation}
Under a gauge transformation \eqref{gaugeXM}, this becomes
\begin{equation}
V(x) \to V(x) \exp\left( -i q \int d^4x' f^\mu(x,x') \partial_\mu \Lambda(x') \right) = V(x) \exp\left( i q \int d^4x' \partial'_\mu f^\mu(x,x' ) \Lambda(x' ) \right)
\end{equation}
and will transform as needed provided
\beq
\partial'_\mu f^\mu(x,x' ) =\delta^4(x - x' )\ .
\eeq
Here we use the requirement that gauge transformations vanish at infinity. Note also that in the quantum theory, one must give a careful definition of the operator $\Phi$, accounting for possible ordering ambiguities.

There is a lot of freedom in choosing the function $f^\mu$.
However not all of this is physical freedom.
Two functions $f^\mu$ are the same if they agree when integrating against all solutions of the equations of motion.
We can think of inequivalent dressings, following \cite{Steinmann1983}, as corresponding to different ``soft photon clouds'' surrounding each particle.
However they need not be composed only of soft modes; for example string-like dressings, which we are about to consider, have divergent energy density.

\subsection{Faraday, or Wilson, line dressing} \label{sec:wils}

Indeed, a particularly simple dressing is that of a spatial Wilson line, or Faraday line, running to the boundary, {\it e.g.} along the $z$ direction:
\beq\label{Wilsz}
\Phi_{W_z}(x)=\phi(x) e^{i q \int_0^\infty ds A_z(x + s \hat z)}\ 
\eeq
where $\hat z$ is the unit vector in the $z$ direction.
Gauge invariance of an operator may be alternately stated as the requirement that the operator creates states satisfying the Gauss' law constraints, and this operator does so by creating an electric field that is localized to an infinitesimally-thin string extending to infinity along the $z$ direction.  

To see that \eqref{Wilsz} creates a line of electric flux, we can use the equal-time commutation relation (see appendix \ref{app:qed}) $[E^i(x), A_j(y)] = i \delta^i_j \delta^{(3)}(\vec x - \vec y)$  to find \cite{Dirac1955} at $t=t'$
\begin{equation} \label{EzWz}
[E^z(x), \Phi_{W_z}(x')] = - q \delta^2(\vec x_\perp - \vec x'_\perp) \theta(z - z') \Phi_{W_z}(x')
\end{equation}
where $x_\perp$ denotes the components orthogonal to the $z$ direction, $x_\perp = (t, x, y)$.  Note that $E_z$ is negative for positive $q$ because $\phi$ creates an antiparticle of charge $-q$.  
The other components of $E$ commute with $\Phi_D$.
The interpretation of this equation is that acting with the operator $\Phi_D$ increases the value of the electric field by the addition of an single line of electric flux.

More explicitly, to find the field created by a general dressed operator $\Phi$, consider 
\beq
A_\mu(x)\Phi(x')|0\rangle= [A_\mu(x),\Phi(x')]|0\rangle + \Phi(x')A_\mu(x)|0\rangle\ .
\eeq
If the commutator is proportional to $\Phi$,
this shows that the field after acting by $\Phi$ differs by ${\tilde A}_\mu(x)$, given by
\beq 
{\tilde A}_\mu(x)\Phi(x') = [A_\mu(x),\Phi(x')] \ ,
\eeq
from that of the vacuum.

To find the future evolution of the field associated to the electric string, we evaluate the commutator to leading order in $q$ in the region that is causally separated from the string, but spacelike to the point $x'$.
In Feynman gauge, this is given by 
\begin{equation} \label{Aline}
\tilde A_z(x) = i q \int_0^\infty ds [A_z(x), A_z(x' + s \hat z)] = \frac{q \, \epsilon(t - t')}{2 \pi \sqrt{-(x_\perp - x'_\perp)^2}},
\end{equation}
where $\epsilon(x)={\rm sign}(x)$.
This corresponds to an electric field which at time $t \neq t'$ is given by
\begin{equation}
\tilde E_z(x) = -\partial_0 \tilde A_z = \frac{q \abs{t - t'}}{2 \pi \left[-(x_\perp - x'_\perp)^2 \right]^{3/2}}\ .
\end{equation} 
This solution describes the field lines of the initial string spreading out.
There is also a corresponding magnetic field, since the electric field is changing with time.

\subsection{Coulomb, or Dirac, dressing}

The string-like configuration considered above is of course a rather unusual electric field for a charged particle; the operator creates a singular electric field concentrated in a single line.  We expect such an operator to receive large corrections in perturbation theory, and this raises serious doubts as to whether such an operator can be rigorously defined in the full quantum theory.  One approach to avoiding this would be to thicken out the field into a finite diameter tube, with a finite stress energy tensor.  A related approach, to which we now turn, is to distribute the field lines even more widely -- providing a more physical dressing for a charged particle.

In particular, a  Coulomb-like field is anticipated to be more typical, and less singular, configuration for the electromagnetic field of a charged particle.
Such a field can be found by {\it averaging} \eqref{Wilsz} over different directions.
Specifically, taking $x=0$, we can average
\beq
\int dz A_z \rightarrow \frac{1}{4\pi}\int d^2\Omega dr\, {\hat r}^i A_i =\int d^3 \vec x \frac{A_r}{4\pi r^2}\ ,
\eeq
which generalizes for $x\neq0$  to give
\beq\label{Ddress}
\Phi_D(x)= \phi(t, \vec x) \exp\left\{i q \int d^3 \vec x'  \frac{(\vec x' -\vec x)^i}{4\pi|\vec x' -\vec x|^3}A_i(t,\vec x' )\right\} \equiv \phi(t,\vec x) V_D(t,\vec x)\ .
\eeq
This is the Dirac\cite{Dirac1955} dressing, which is sometimes rewritten after an integration by parts,
\beq\label{Dsurf}
\int d^3 \vec x \frac{A_r}{4\pi r^2} = \int d^3 \vec x \frac{1}{4\pi r} \partial^i A_i - \int \frac{d^2\Omega}{4\pi} x^i A_i(\infty)   \ .
\eeq
Using the canonical commutators, one may check that this creates a Coulomb field at time $t$:
\beq
[E^i(x),\Phi_D(x')]_{\big|_ {t=t'}}= -\frac{q}{4\pi} \frac{(\vec x - \vec x')^i}{|\vec x - \vec x'|^3}\Phi_D(x)\ .
\eeq

For earlier or later times we can use the fact that the commutator $[A_\mu(x),\Phi(x')]$ satisfies the Heisenberg equations of motion, with initial data given by the equal-time commutators.  Outside the lightcone of $x'$, these are, working to linear order in $q$, the source-free Feynman-gauge equations $\square  [A_\mu(x),\Phi_D(x')] = 0$.
The gauge-dependent term $\partial^\mu A_\mu$ does not appear in this equation, since $\partial^\mu A_\mu$ generates gauge transformations (see Appendix \ref{app:qed}) and so
\begin{equation}
 [\partial^\mu A_\mu(x), \Phi_D(x')] = 0.
\end{equation}
This also means the solution $\tilde A$ satisfies the Lorenz gauge condition $\partial^\mu {\tilde A}_\mu=0$.  
The free field equation with these initial conditions has solution
\beq\label{gcoul}
\tilde A_i(x) = -\frac{q}{4\pi}(t-t')\frac{(\vec x' - \vec x)_i}{\abs{\vec x' - \vec x}^{3}} \quad,\quad {\tilde A}_0=0\ ,
\eeq
giving the static electric field of a particle at $\vec x'$. 

The field configuration \eqref{gcoul} may be continued into the future lightcone of $x'$, but then does not in general satisfy the equations of motion.  The reason is that the solution there depends on the state of motion of the particle at and after $t'$.  In general, this solution will have, in addition to \eqref{gcoul}, a radiative part.  The full determination of this field depends on  particle motion in the interacting theory.

As we have noted, if 
the Wilson line operator $\Phi_{W_z}$ acts on the vacuum, the highly-localized electric field it creates is not stable, and will decay \cite{Shab,HaJo}.  Another way to see its relation to the Dirac field is  to rewrite the field arising from the Wilson line  as a longitudinal, Coulomb, piece, plus a transverse, radiative, piece\cite{BLMdecomp}, using
\beq\label{raddecomp}
A_i = A_i^L + A_i^T = \frac{\partial_i\partial_j}{\nabla^2}A_j + \left(\delta_{ij} - \frac{\partial_i\partial_j}{\nabla^2}\right)A_j\ .
\eeq 
The integral $\int ds A_z^L$ then gives the Dirac dressing  \eqref{Dsurf}; $\int ds A_z^T$ contributes extra field energy which radiates to infinity. 

\subsection{Worldline dressing}\label{WLQED}

An electromagnetic dressing that takes particle motion into account is defined by the world-line expression
\beq\label{WLdress}
\Phi_{WL}(x) = V_D(a,\vec 0) {\tilde T} \exp\left\{-iq\int_{0}^{1} d\lambda \frac{dx^\mu}{d\lambda}A_\mu(x^\nu(\lambda))\right\}\phi(x)\ .
\eeq
Here $V_D$ is the Dirac dressing given in \eqref{Ddress}, $\tilde T$ anti-orders in time (latest time is to the right), and 
$x^\mu(\lambda)$ is a trajectory with   $x^\mu(0) = (a,\vec 0)$ and $x^\mu(1)=x^\mu$.   This produces the electromagnetic field that results from classical motion along $x(\lambda)$.  
For large $m$, a quantum particle approximately follows such a trajectory, and \eqref{WLdress} produces the leading-order field, but for general $m$ there are quantum corrections to this field.  

To see that \eqref{WLdress} produces the field of the moving particle, we consider again the commutator $\tilde A_\mu$.
This will contain a term from $V_D$ as well as one from the worldline, given by
\begin{equation}
\left[ A_\mu(x'), -iq\int_{0}^{1} d\lambda \frac{dx^\nu}{d\lambda}A_\nu(x^\rho(\lambda)) \right] = q \int d\lambda D_{\mu \nu}(x' - x(\lambda)) \frac{d x^\nu}{d \lambda} + \calo(q^2)\ ,
\end{equation}
where $D_{\mu\nu}$ is the Pauli-Jordan function for $A_\mu$, \eqref{EMprop}.
When $x'$ is spacelike to $x$ and to the future of $(a,\vec 0)$, $D_{\mu \nu}$ agrees with the retarded propagator so this term produces the field of the moving particle.
When $x'$ is spacelike to $x$ and to $(a, \vec 0)$, we have the Coulomb field \eqref{gcoul}.

We note that although the vector potential $A_\mu$ is discontinuous across the lightcone of $(a, \vec 0)$, this is a pure gauge discontinuity and $F_{\mu \nu}$ will be continuous provided the trajectory $x(\lambda)$ is initially at rest. 
If $\tilde A_\mu$ is initially at rest, then the field just to the future of the lightcone of $(a, \vec 0)$ is 
\begin{equation} \label{Amu-worldline}
\tilde A_\mu(x') = -\eta_{\mu 0} \frac{q}{4 \pi \abs{\vec x - \vec x'}}\ ,
\end{equation}
which is related to the configuration \eqref{gcoul} (with $x\leftrightarrow x'$) by a gauge transformation \begin{equation}
\Lambda(x') = \frac{q (t - t')}{4 \pi \abs{x - x'}}\ .
\end{equation}

Let us examine the large-$m$ case more closely.  In the large-$m$ limit, the four-velocity $u^\mu$ of a particle is constant, and superselected\cite{Georgi1990}.  Then, from the equation of motion
\beq
D^\mu D_\mu \phi=m^2\phi\ ,
\eeq 
one may show
\beq \label{dressing-eq-qed}
i u^\mu \partial_\mu \Phi_{WL}(x)= m \Phi_{WL} - e^{i m u^\mu x_\mu} \frac{D^2}{2m} \left(e^{-i m u^\mu x_\mu}\phi\right)e^{iq\int_0^1A} V_D\ .
\eeq 
Thus, to leading order in $1/m$, $\Phi_{WL}$ creates a state that is an energy eigenstate; this statement was called the ``dressing equation" in \cite{Bagan1999a}.  

The latter feature, together with the fact that in the asymptotic regime, where particles have large separation, $m$ is the largest energy scale, motivated \cite{Bagan1999a,Bagan1999b} to use \eqref{WLdress}, with constant-velocity trajectories, to define asymptotic states which avoid the usual infrared difficulties of  describing asymptotic particle states in QED.

\subsection{Lorenz dressing}

One natural way to specify a gauge-invariant operator is to fix a gauge by some condition, and define $\Phi(x)$ to be $\phi(x)$ in that particular gauge.
If this condition completely fixes the gauge, we can solve for the gauge transformation $\lambda$ that transforms an arbitrary vector potential $A_\mu$ into this chosen gauge.
This allows us to express $\Phi(x)$ in a manifestly gauge-invariant way,
\begin{equation}
\Phi(x) = e^{-iq \lambda(x)} \phi(x)
\end{equation}
where $\lambda(x)$ is a nonlocal function of $A_\mu$.  Indeed, the preceding construction of $\Phi_{W_z}$, for example, gives the field operator in axial gauge.

Another apparently natural choice for this purpose is Lorenz gauge $\partial_\mu A^\mu = 0$.
However this is not a complete specification of the gauge: given a vector potential $A_\mu$, we must find the gauge transformation $\lambda$ to Lorenz gauge, which is found by solving
\begin{equation}
\partial^\mu A_\mu = \square \lambda.
\end{equation}
The solutions to this equation are not unique until we specify appropriate initial (or final) data for ${\lambda}$.

If our initial data surface is spacelike, we should choose initial data that determine both $\lambda$ and its first derivative away from the surface.
For example, we can fix $n^\mu A_\mu = 0$, where $n^\mu$ is a normal to an initial data surface, which we take to be a spacelike of constant time.
This leads to an initial condition for $n^\mu \partial_\mu \lambda = n^\mu A_\mu$.
This does not yet determine a solution for $\lambda$, since we need to know both the initial value of $\lambda$ and its time derivative.
We can give an additional constraint $\partial^i A_i = 0$, which leads to the equation $\vec \nabla^2 \lambda = \partial^i A_i$ on the initial surface.
This determines $\lambda$ completely, once we include the boundary condition $\lambda \to 0$ at spacelike infinity.

Of course there are other ways to fix the gauge asymptotically.
Since $\lambda$ satisfies a free wave equation, it may make more sense to fix its initial data on $\scri^-$.
For example, we can impose $A_u = 0$, where $u$ is the null generator of $\scri^-$.
This leads to the equation $\partial_u \lambda = A_u$, which then determines $\lambda$ completely on $\scri$ from the condition that the gauge transformations vanish at infinity, $\lambda = 0$ on $i^0$.

One can therefore likewise formulate dressings based on these gauges; we leave description of these dressings to future work.

\section{Gauge-invariant observables for gravity}\label{Gravobs}

We next turn to a discussion of gravitational observables analogous to those we have described in QED.  The nonlinearity of gravity of course is a significant complication, which we manage by working perturbatively in Newton's constant $G_D$. In particular, the leading-order, linear structure of gravity has important similarities to that of QED.  This structure will describe the dominant behavior in the weak-field regime.

Specifically, consider the gravitational Lagrangian, minimally coupled to a scalar $\phi$ with mass $m$,
\beq 
\call= \frac{2}{\kappa^2}R -\frac{1}{\alpha\kappa^2} \frac{\sqrt{\abs {g^0}}}{\sqrt{\abs g}}\frac{1}{|g|}\left[\nabla^0_\mu\left(\sqrt{|g|} g^{\mu\nu}\right)\right]^2 -\hf\left[(\nabla\phi)^2+m^2\phi^2\right]
\eeq
where $\kappa^2=32\pi G_D$, and $R$ is the scalar curvature.  The derivative $\nabla^0$ is with respect to the background metric about which we perturb.  The parameter $\alpha$ is a gauge fixing (more precisely, breaking) parameter; $\alpha=0$ gives de Donder gauge, an analog of Lorenz or Landau gauge, $\alpha=1$ is an analog of Feynman gauge for QED, and $\alpha \to \infty$ is the analog of unitary gauge, which restores the gauge symmetry.  The perturbative expansion follows from
\beq
g_{\mu\nu}=\eta_{\mu\nu} +\kappa h_{\mu\nu}\ ,
\eeq
where here the background is taken to be the Minkowski metric.  Further details of the perturbative theory are supplied in appendix \ref{app:gravity}.

As was the case for a charged field in QED, the field $\phi(x)$ is not a gauge-invariant operator, since under a diffeomorphism $f: M \to M$, the field transforms as the pushforward
\beq
\phi \to f_* \phi \  , \qquad (f_* \phi)(x) = \phi(f^{-1}(x))
\eeq
If we take $f$ to be an infinitesimal diffeomorphism, $f^\mu(x) = x^\mu + \kappa \xi^\mu$,
\beq \label{deltaphi}
\delta \phi(x) = \phi(f^{-1}(x)) - \phi(x) = - \kappa \xi^\mu \partial_\mu \phi\ + O(\kappa^2).
\eeq
To find a diffeomorphism-invariant operator, we will form a composite operator that includes both $\phi$ and the gravitational field
sourced by $\phi$.

A simple approach to constructing operators invariant under linearized diffeomorphisms, analogous to the approach taken for QED, is to seek an operator of the form
\beq
\Phi(x) = e^{iV^\mu(x) P_\mu}\phi(x)e^{-iV^\mu(x) P_\mu}=\phi(x^\mu+V^\mu(x)) = \phi(x) + V^\mu(x) \partial_\mu \phi(x) + \calo(V^2)\ ,
\eeq
where $P_\mu=-i\partial_\mu$ (compare the QED expression \eqref{Fieldxm}).  Here the ``dressing" $V^\mu(x)$ is a functional of the metric perturbation that transforms under a diffeomorphism as
\begin{equation}
x^\mu + V^\mu(x) \to f^\mu(x + V(x))\ ,
\end{equation}
which at linear order becomes
\beq \label{Vtrans}
\delta V^\mu(x) = 
\kappa\xi^\mu(x)\ .
\eeq 
Then, using the combined transformations,  at linear order
\begin{equation}\label{Phitrans}
\delta \Phi(x) = \delta[\phi(x+V)] 
\approx \delta \phi(x) + \delta V^\mu(x) \partial_\mu \phi(x)
= 0\ .
\end{equation}
The transformation law \eqref{Vtrans} 
should follow from the change in the  metric, which transforms as 
\beq
\delta g_{\mu \nu} = - \kappa \mathcal{L}_\xi g_{\mu \nu} = - \kappa ( \nabla_\mu \xi_\nu + \nabla_\nu \xi_\mu)\ ;
\eeq
correspondingly the metric perturbation transforms as
\begin{equation} \label{lindiff}
\delta h_{\mu \nu} = -\partial_\mu \xi_\nu - \partial_\nu \xi_\nu + O(\kappa)\ .
\end{equation}
This paper will primarily (though not exclusively) consider such constructions at leading order in $\kappa$, and in so doing will consider only linearized diffeomorphisms and drop the $O(\kappa)$ term from \eqref{lindiff}.  A construction to higher-order in $\kappa$ will then be constrained by this leading behavior; we save general examination of all-orders behavior for future work.

As in the case of gauge theory, we can try to find $V$'s satisfying \eqref{Vtrans} that are of the form
\begin{equation} \label{linear}
V^\mu(x) = \kappa \int d^4 x' \; f^{\mu \nu \lambda}(x, x') h_{\nu\lambda}(x')\ ,
\end{equation}
where $f$ is assumed symmetric in its last two indices $f^{\mu \nu \lambda} = f^{\mu \lambda \nu}$. Given \eqref{lindiff}, in order for $V^\mu$ to transform as \eqref{Vtrans}, the function $f$ parameterizing the dressing must satisfy
\begin{equation}
2 \partial_\nu' f^{\mu \nu \lambda} (x,x') = \delta^{4}(x-x') \eta^{\mu \lambda}\ .
\end{equation}
Again, there is substantial freedom in choosing the functions $f$, which determine the ``soft graviton cloud" (plus, possibly, a harder component) of a $\phi$ particle, though not all freedom yields physically-inequivalent dressings.

\subsection{Gravitational Wilson line} \label{sec:gwils}

As in QED, a very simple dressing is a Wilson line. One very geometrical way to think of such a line in gravity arises if points are located by shooting a spatial geodesic in a perpendicular direction from a fixed asymptotic ``platform," where diffeomorphisms are taken to vanish; ultimately, we might take this to reside at infinity.\footnote{For an analogous construction in anti de Sitter space, with the AdS boundary providing a platform, see \cite{Heemskerk2012,Kabat2013,Almheiri2014}.}  Specifically, the location of a general point is determined by the platform point at which the geodesic originates, and the distance along the geodesic. The field $\phi$, expressed as a function of this data, is gauge invariant under gauge transformations vanishing at the platform. 

We can thus choose as coordinates the initial platform position, and the geodesic distance.  Specifically, define new coordinates $\breve x^\mu$ in which the $\breve z$ direction is perpendicular to the platform, and choose the ``axial" gauge (or, geodesic normal coordinates)
\beq\label{axmet}
h_{\breve z \breve \mu}=0\ .
\eeq 
In this gauge, the $\breve z$ coordinate is the geodesic distance, and the remaining coordinates correspond to the platform position.  The diffeomorphism-invariant field is thus just the scalar field $\phi$ in these coordinates.
To write this in terms of the scalar field in arbitrary coordinates, consider a new (general) coordinate system 
\beq\label{brevtox}
x^{\mu}= f^\mu(\breve x) \ . 
\eeq
Note that, for general coordinates $x^{\mu}$, the function $f^\mu$ depends nonlocally on the metric, since it must take a general metric to the form \eqref{axmet}.
Then, 
\beq \label{gwilsdef}
\Phi_{W_z}(x) = \phi(f(x))
\eeq
is the diffeomorphism-invariant field written in terms of the field $\phi$ expressed  in general coordinates.  (At this stage we drop the accent on $\breve x$.)

The preceding statements are valid to all orders in the perturbative expansion in $\kappa$, but can also be simply explained to leading order in this expansion.  Specifically, consider a small metric perturbation and parameterize the relation between axial coordinates $\breve x^\mu$ and a general coordinate $x^{\mu}$ as
\beq \label{axrel}
\breve x^\mu = x^{\mu} - V_{W_z}^{\mu}\ .
\eeq
Then, the relation \eqref{gwilsdef} becomes
\begin{equation}\label{gwilsdefl}
\Phi_{W_z}( x) \approx \phi( x^\mu+ V_{W_z}^{\mu}( x))
\end{equation}
to leading order.
Given the linearized gauge transformation \eqref{lindiff}, the axial gauge \eqref{axmet} can be fixed by a diffeomorphism of the form
\beq \label{WilsV}
-\kappa\xi_z =\frac{\kappa}{2} \int_0^\infty ds\, h_{zz}(x+ s \hat z)= V_{W_z,z}(x)\quad \;\quad -\kappa \xi_{\check \mu}  = \kappa \int_0^\infty ds\left[h_{z \check \mu}(x+s \hat z) +\hf\partial_{\check \mu} \int_{s}^\infty ds'\, h_{zz}(x+s' \hat z)\right]=V_{W_z,\check \mu}(x)\ 
\eeq
where $\check \mu$ denotes indices excluding $z$, and the platform has been taken to $z=\infty$.  

With these expressions, one can explicitly check the leading-order diffeomorphism invariance of \eqref{gwilsdef}, \eqref{gwilsdefl}.  From the metric transformation \eqref{lindiff}, $V_{W_z}$ transforms as in \eqref{Vtrans} for a diffeomorphism $\xi$ that vanishes at infinity, so according to \eqref{Phitrans}, $\Phi_{W_z}$ is diffeomorphism invariant.
The gauge-invariant operator $\Phi_{W_z}$ is the gravitational analog of  the electric Wilson line operator defined in \eqref{Wilsz}.

This construction can also be characterized by solving the geodesic equation for a curve $x^\mu=\breve x^\mu+s{\hat z}^\mu+v^\mu(s)$ with the boundary condition $v^\mu(\infty)=0$.
The quantity $V^\mu(x)=v^\mu(0)$ then gives the coordinate displacement between general coordinates and axial coordinates.
To find the linearized dressing $V^\mu$, we solve the linearized geodesic equation: 
\begin{equation}
\partial_s^2 v^\mu(s) + \Gamma^\mu_{zz}(x^\mu + s \hat z^\mu) = 0 
\end{equation}
which gives
\beq\label{WilsC}
V^\mu_{W_z}(x)= - \int_0^\infty ds \; s \; \Gamma^\mu_{zz}(x+s{\hat z}) +{\rm s.t.}\ ,
\eeq
where s.t. denotes  a surface term at infinity.\footnote{This vanishes if  $|\vec x|h_{\mu z}(x)$ vanishes in the limit $\vec x\rightarrow\infty$.}  This can be checked to be equivalent to \eqref{WilsV}.

As with the case of electromagnetism, the operator \eqref{gwilsdef}, \eqref{gwilsdefl} creates a singular metric configuration, where gravitational field lines are concentrated in an infinitesimal string.  
This can be seen by studying the commutator of $h_{\mu\nu}$ with $V_{W_z}$.  
Even in the linear theory, this is an unstable field configuration, and if it is created the field lines will then dynamically spread out into a Coulomb-like configuration.
Of course in the non-linear theory we also expect such a concentrated gravitational field to source large non-linear corrections,  and as a result at a minimum the gravitational field configuration should be ``thickened\cite{SGalg};"
 the absence of string-like solutions to the sourceless Einstein equations moreover likewise strongly indicates that the field configuration should undergo
 such a decay to a more symmetric configuration in the full theory.

\subsection{Gravitational Coulomb dressing}\label{GravCoul}

\subsubsection{Construction}

Like with QED, an approach to finding a more symmetric dressing is to average the gravitational Wilson line over all directions.
This is most easily done starting with \eqref{WilsC}, which then becomes, specializing to $x=0$,  and working in $D=4$, 
\beq\label{vmuaverage}
V^\mu_C(0)=
 -\frac{1}{4 \pi} \int d^2 \Omega \int_0^\infty dr \; r \; \Gamma^\mu_{\alpha \beta} \hat r^\alpha \hat r^\beta +{\rm s.t.}
= -\frac{1}{4 \pi} \int d^3 x\frac{1}{r} \; \Gamma^\mu_{\alpha \beta} \hat r^\alpha \hat r^\beta
 +{\rm s.t.}
\eeq
with $\hat r$ the unit radial vector.  While we use $D=4$ for much of the following, the discussion readily generalizes to $D>4$.
Specializing to the timelike component, we find
\begin{equation}\label{GC0}
V_C^0(0) = -\frac{\kappa}{4 \pi} \int d^3x \left( \frac{\partial_0 h_{rr}}{2 r} + \frac{h_{0r}}{r^2} \right)
\end{equation}
where the surface term has cancelled.
The spatial components become
\begin{equation}\label{GCi}
V_C^i(0) = \frac{3 \kappa}{8 \pi} \int d^3x \frac{\hat r^i}{r^2} h_{rr},
\end{equation}
where again the surface terms cancel. 
At a general location $x$, these become
\beq
V_C^0(x) = -\frac{\kappa}{4 \pi} \int d^3x' \left( \frac{\hat d^\alpha\hat d^\beta \partial_0 h_{\alpha\beta}}{2 d} + \frac{h_{0\alpha}\hat d^\alpha}{d^2} \right)\quad ,\quad V^i_C(x)=\frac{3 \kappa}{8 \pi} \int d^3x' \frac{\hat d^i\hat d^\alpha\hat d^\beta }{d^2} h_{\alpha\beta}\ 
\eeq
where  $\vec d = \vec x'-\vec x$.
One can check directly that $V^\mu_C$ has the correct gauge variation; using  \eqref{lindiff}, one finds 
\begin{eqnarray}
\delta V_C^i(x) &=& \kappa \xi^i(x) - \frac{3 \kappa}{4 \pi} \int d^2\Omega' \frac{r^{\prime 2}}{d^2} \hat r'\cdot \hat d\ \hat d\cdot\xi \ \hat d^i \\
\delta V_C^0(x) &=& \kappa \xi^0(x) + \frac{\kappa}{4 \pi} \int d^2\Omega' r^{\prime 2} \hat r'\cdot \hat d\left(\frac{\hat d\cdot \dot\xi}{d}+\frac{\xi_0}{d^2}\right)\ .
\end{eqnarray}
Thus we recover the expected transformation law, $\delta V_C^\mu(x) = \kappa \xi^\mu(x)$, as in \eqref{Vtrans},  for $\xi$ satisfying falloff conditions such that the boundary terms in these expressions vanish, i.e. $\xi_r$ and $r \dot \xi_r + \xi_0$ vanish as $r \to \infty$.

\subsubsection{Dressing field}\label{Dressfield}

Comparing QED, we might expect that the operator $\Phi_C(x)=\phi(x+V_C(x))$ creates a gravitational analog of the Coulomb field.  In order to check this, we examine the commutator $[h_{\mu\nu}(x),\Phi_C(0)]$, which indicates, as with the QED case, how the gravitational field is changed by action of the dressed operator.

As a first step, we need the commutator
\begin{align}
[\bar h_{\mu \nu}(x'), V^0_C(0)] 
&= -\frac{\kappa}{4 \pi} \int d^3x \left( \frac{\hat r^{\lambda} \hat r^{\sigma}}{2 r} \partial_0 + \frac{\hat t^\lambda \hat r^\sigma}{r^2} \right) [\bar h_{\mu \nu}(x'), h_{\lambda \sigma}(x)] \nonumber \\
&= -\frac{i\kappa}{4 \pi}  \int d^3x \left( \frac{\hat r_\mu \hat r_\nu}{2 r} \partial_0 + \frac{\hat t_{(\mu} \hat r_{\nu)}}{r^2} \right) D(x'-x)\ ,
\end{align}
where we work in Feynman gauge and $D$ is the Pauli-Jordan propagator of a massless scalar, \eqref{Dscalar}.
This can be rewritten in terms of scalar integrals using (see appendix \ref{formulas})
\begin{equation}
\frac{\hat r_\mu \hat r_\nu}{r} = \frac{q_{\mu \nu}}{r} - \partial_\mu \partial_\nu r, \qquad \frac{\hat r_\nu}{r^2} = - \partial_\nu \frac{1}{r}\ ,
\end{equation}
where the spatial metric is denoted $q_{\mu\nu}$.  We then integrate by parts, so that the derivatives act on $D(x'-x)$, and then trade them for $x'$ derivatives, which can be pulled outside the integral:
\begin{align} \label{barh}
[\bar h_{\mu \nu}(x'), V_C^0(0)] 
&=  \frac{i\kappa}{4 \pi} \int d^3 x \left( \hf\partial_\mu \partial_\nu r \partial_0 - \frac{q_{\mu \nu}}{2r} \partial_0 + \hat t_{(\mu} \partial_{\nu)} \frac{1}{r} \right) D(x'-x) \nonumber \\
&= -\frac{i\kappa }{4 \pi} \vec \partial'_\mu \vec \partial'_\nu \partial'_t \left( \int d^3x \; \frac{r}{2} D(x'-x) \right) + \frac{i\kappa}{4 \pi} \left( \frac{q_{\mu \nu}}{2} \partial'_0 + \hat t_{(\mu} \vec \partial'_{\nu)} \right) \left( \int d^3x \frac{D(x'-x)}{r} \right)\ ,
\end{align}
{where $\vec \partial$ denotes the spatial gradient.}
The resulting scalar integrals can then be evaluated:
\begin{align}
\int d^3x \; \frac{1}{r} \; D(x'-x) &=
\begin{cases} 
-1 &\mbox{if } t' > r' \\ 
-t'/r' &\mbox{if } -r' < t' < r' \\
1 &\mbox{if } t' < -r' 
\end{cases}
\end{align}
\begin{align}
\int d^3x \; r \; D(x'-x) =
\begin{cases} 
-t^{\prime2} - \frac{r^{\prime2}}{3} &\mbox{if } t' > r' \\ 
-t'r' - \frac{t^{\prime3}}{3 r'}. &\mbox{if } -r' < t' < r' \\
t^{\prime 2} + \frac{r^{\prime2}}{3} &\mbox{if } t' < -r' \ .
\end{cases}
\end{align}
Using these result with \eqref{barh} and combining terms we find, in the region that is spacelike to the origin, $-r < t < r$,
\beq\label{V0field}
[\bar h_{\mu \nu}(x), V_C^0(0)]  
= -\frac{i \kappa}{4 \pi} \left[ \frac{\hat r_\mu \hat r_\nu}{2 r} {-} t \frac{ \hat t_{(\mu} \hat r_{\nu)}}{r^2} + t^2 \frac{q_{\mu \nu} - 3 \hat r_\mu \hat r_\nu}{2 r^3} \right].
\eeq
In the regions that are timelike to the origin, $t > r$ and $t < -r$, we find $[\bar h_{\mu \nu}(x), V_C^0(0)] = 0$.
We can check explicitly that \eqref{V0field} satisfies the harmonic gauge condition $\partial^\mu \bar h_{\mu \nu} = 0$ as well as the equation of motion $\square \bar h_{\mu \nu} = 0$. 

We also need the commutator 
$[\bar h_{\mu \nu}(x), V^i_C(0)]$.
Only the spatial components contribute, and they are given by 
\begin{equation}
[\bar h_{jk}(x'), V_C^i(0)] = \frac{3i\kappa}{8 \pi} \int d^3x \frac{\hat r^i \hat r_j \hat r_k}{r^2} D(x'-x).
\end{equation}
We use the same trick of expressing the tensor as a derivative (see \eqref{3deriv}),
\begin{equation}
\frac{\hat r_i \hat r_j \hat r_k}{r^2} = \frac{1}{3} \partial_i \partial_j \partial_k r - q_{(ij} \partial_{k)} \frac{1}{r}\ ,
\end{equation}
integrate by parts, pull the derivatives outside the integral, and find
\beq\label{Vifield}
[\bar h_{jk}(x), V_C^i(0)] = -\frac{3i\kappa}{8 \pi} \left( \frac{t}{r^2} \hat r_i \hat r_j \hat r_k + \frac{t^3}{9} \partial_i \partial_j \partial_k \frac{1}{r} \right)
\eeq
in the case where $x$ is spacelike to the origin.  In the timelike regions $t > r$ and $t < -r$ we have $[\bar h_{\mu \nu}(x), V_C^i(0)] = 0$.
We can again check that the result \eqref{Vifield} satisfies the harmonic gauge condition $\partial^\mu \bar h_{\mu \nu} = 0$ and the equation of motion $\square \bar h_{\mu \nu} = 0$.

We first consider the dressing field due to a massive particle, which we can approximately consider to be at rest, so that $\partial_0 \phi(x) =  i m \phi(x)$ (for the creation part of the operator) and we can neglect  spatial gradients.
Then we have
\begin{equation} \label{schw-nonstatic}
[\bar h_{\mu \nu}(x), \Phi{(0)}] = [\bar h_{\mu \nu}{(x)}, V_C^0{(0)}] \partial_0 \phi{(0)} = \frac{\kappa m}{4 \pi} \left[ \frac{\hat r_\mu \hat r_\nu}{2 r} {-} t \frac{ \hat t_{(\mu} \hat r_{\nu)}}{r^2} + \frac{t^2}{2 r^3} (q_{\mu \nu} - 3 \hat r_\mu \hat r_\nu) \right] \phi(0) = \tilde {\bar h}_{\mu \nu}(x) \phi(0)\ ,
\end{equation}
which gives the field of the particle,  valid for all points spacelike to the origin, $\abs{t} < r$.
As noted, this  satisfies the harmonic gauge condition and the equation of motion $\square \tilde h_{\mu \nu} = 0$ in this region.

While the metric in \eqref{schw-nonstatic} may be unfamiliar, it is simply the linearized Schwarzschild metric in an unusual gauge.
By means of an infinitesimal diffeomorphism
\begin{equation}
\xi_\mu = -\frac{ \kappa m}{16 \pi} (\tfrac12 - t^2/r^2) \hat r_\mu,
\end{equation}
we can put the metric into the form:
\begin{equation}
h_{\mu \nu} - \partial_\mu \xi_\nu - \partial_\nu \xi_\mu = \frac{\kappa m}{16 \pi r} (\hat t_\mu \hat t_\nu + \hat r_\mu \hat r_\nu)
\end{equation}
which is the linearized  Schwarzschild solution.  This can also be checked by calculating the ADM energy,
\begin{equation}
P^0_\text{ADM} = \frac{1}{16 \pi G} \oint r^2 d^2 \Omega \; \hat r^i (\partial_j g_{ij} - \partial_i g_{jj})  = \frac{2}{\kappa} \oint r^2 d^2 \Omega \; \hat r^i (\partial_j h_{ij} - \partial_i h_{jj})\ .
\end{equation}
which indeed yields $P^0_{\rm ADM}=m$ for the solution in \eqref{schw-nonstatic}.

Generalizing to the case of a localized source with $\partial_\mu \phi \approx -ip_\mu \phi$, we also pick up a field contribution from \eqref{Vifield}, of the form
\beq \label{hfromVi}
\bar h_{jk}= -\frac{3 \kappa}{8\pi} p^i \left( \frac{t}{r^2} \hat r_i \hat r_j \hat r_k + \frac{t^3}{9} \partial_i \partial_j \partial_k \frac{1}{r} \right)\ .
\eeq
Again, this satisfies  the harmonic gauge condition and the equation of motion $\square h_{\mu \nu} = 0$ in the region spacelike to the origin.   

As a check, we can calculate the ADM momentum of the solution.
This is given by \cite{Regge1974}
\begin{equation}
P^i_\text{ADM} 
= - \frac{2}{\kappa} \oint r^2 d^2\Omega \; \hat r_j \pi^{ij}_\text{ADM} 
= -\frac{2}{\kappa} \oint r^2 d^2 \Omega \; \hat r^j \left( \dot h_{ij} + \partial_i h_{0j} - \partial_j h_{0i} - \dot h_{kk} \delta_{ij} \right)\ ,
\end{equation}
where $\pi^{ij}_\text{ADM}$ is the linearized canonical momentum in the ADM formalism:
\begin{equation}
\pi^{ij}_{\text{ADM}} = \dot h_{ij} - \partial_i h_{0j} - \partial_j h_{0i} - \dot h_{kk} \delta_{ij} + 2 \partial_k h_{0k} \delta_{ij}\ ,
\end{equation}
and we have used the identity $\oint r^2 d^2 \Omega \; \hat r^j (\partial_k h_{0k} \delta_{ij}) = \oint r^2 d^2 \Omega \; \hat r^j (\partial_i h_{0j})$.
We can verify that the solution \eqref{hfromVi} has $P_\text{ADM}^i = p^i$, as expected.
This is a general consequence of the commutation relations
\begin{equation} \label{Pcommutator}
[P_\text{ADM}^\mu, V_C^\nu(x)] = i \eta^{\mu \nu}\ .
\end{equation}

Another check on the preceding constructions is to consider the commutator 
\begin{equation}
[P^\mu_\text{ADM}, \Phi(x)] = i \partial^\mu \phi_{\big\vert_{x+V}}= i\partial^\mu\Phi(x)+\calo(\kappa)\ .
\end{equation}
This shows that the ADM momentum generates translations of the diffeomorphism-invariant observables.
Note, however, that at this order $P_\text{ADM}$ only generates translations of the field $\phi$ and not of the dressing; this is a consequence of truncating our perturbation theory at $\calo(\kappa)$.
Since $P_\text{ADM}$ contains an explicit factor of $\kappa^{-1}$, one has to go to order $\kappa^{n+1}$ to see the effect of translating $V$ in the operator $\Phi$ at order $\kappa^{n}$.
We expect that the ADM $D$-momentum will continue to generate translations of $\Phi$ at higher orders in perturbation theory.

\subsubsection{Further comments on different dressings}

We have noted above that the gravitational Wilson line will in general decay to a Coulomb-like configuration.  As with QED, the averaging procedure yielding  
\eqref{vmuaverage} projects the Wilson-line field onto this configuration, and the remaining transverse field, as in \eqref{raddecomp}, represents the radiative part of the gravitational field.

Other simple examples of different dressings satisfying the constraints can be obtained by adding to a $V^\mu$ giving a solution of  the constraints a multiple of the deformations
\beq\label{deltav0}
\Delta V^0(x)= \frac{\kappa}{8\pi} \int d^3x' \frac{\hat d ^i \hat d^j + q^{ij}}{d} \dot h_{ij}
\eeq
or 
\beq\label{deltavi}
\Delta V^i(x) = \frac{\kappa}{8\pi} \int d^3x' \frac{3\hat d ^i \hat d^j \hat d^k + \hat d^i q^{jk} - 2 q^{ij} \hat d^k}{d^2} h_{jk}
\eeq
again with $\vec d=\vec x'-\vec x$.  These can be shown to be invariant under diffeomorphisms \eqref{lindiff} which vanish sufficiently rapidly at infinity, and thus 
adding them maintains the condition \eqref{Vtrans}; moreover they shift the dressing field by a diffeomorphism.
Note that a linear combination of them can be added to the Coulomb dressing to cancel the leading $\xi$ dependence at infinity, resulting in invariance under supertranslations at spatial infinity \cite{Ashtekar1978,Ashtekar1992}.

\subsection{Gravitational worldline dressing}\label{GravWL}

In parallel with QED, the dressing
\eqref{vmuaverage} creates the field for a particle at $x$ that has been at rest forever.  Note also that it does not necessarily give the correct field in the future light cone of the point $x$, since that will depend on the subsequent state of motion of the particle.  We can  consider gravitational dressings corresponding to more general trajectories for the particle; for example, for an operator describing a particle at $x$, we might want to consider gravitational dressings corresponding to different paths that the particle took to $x$.  For completeness, we give a construction relevant to these cases.
A possible way to account for this is to construct a dressing similar to that of \eqref{WLdress}, beginning with the Coulomb dressing \eqref{vmuaverage} at a time in the distant past, and then adding a worldline component to account for the motion of the particle up to a given time. In principle we might try to consider an arbitrary worldline, specified in an invariant way, {\it e.g.} by specifying its acceleration in a local frame carried by parallel transport.  However, this will not yield a dressing creating a field solely generated by a particle following that worldline, since the gravitational field must be coupled to a conserved stress tensor, which for an isolated particle must correspond to a geodesic.  Thus allowed worldlines are  determined by the geometry, and the final position and momentum of the particle.

Alternatively, we can proceed directly to construction of the worldline-dressed operator directly analogous to \eqref{gwilsdefl}, via a geodesic construction.  Specifically, we define a worldline-dressed operator of the form
\begin{equation}\label{WLdressg}
\Phi_{WL}(x)\approx\phi(x^\mu+ V^\mu_{WL}(x,x') + V^\mu_C(x'))
\end{equation}
where the $\approx$ denotes that we are again working only to linear order in the metric perturbation.  
Here $V^\mu_C(x')$ dresses the point $x'=(t', {\vec x})$ {directly to the past of $x = (t,\vec x)$}, and transforms as $\delta V^\mu_C(x')=\kappa\xi^\mu(x')$, as before.  This is the special case of a geodesic with zero initial velocity, and will be generalized below.
Then the worldline part of the dressing can be associated with the choice of a gaussian-normal gauge $h_{0\mu}=0$ for $t>t_0$, analogous to the axial gauge choice.  Effectively, one localizes spatial points at $t=t'$ with respect to the boundary, and then localizes future points by relative to these.
As in eq.~\eqref{WilsV}, this gives an expression 
\begin{eqnarray} \label{WLV}
V_{WL,0}(x,x') &=&-\frac{\kappa}{2} \int_0^{t-t'} d\lambda\, h_{00}(x' + \lambda \hat t)\quad \ ;\cr 
V_{WL,i}(x,x') &=& - \int_0^{t-t'} d\lambda\left[\kappa h_{0i}(x' + \lambda \hat t) -\hf\partial_i \int_{0}^\lambda d\lambda'\, \kappa h_{00}(x' + \lambda' \hat t) + \partial_iV_{C,0}(x')\right]\ .
\end{eqnarray}
The latter term in $V_{WL}^i$ is needed so that
\beq\label{VWLvar}
\delta V_{WL}^\mu(x,x') = \kappa \xi^\mu(x) - \kappa \xi^\mu(x')\ ,
\eeq
as is readily checked.
Then, under a general diffeomorphism
\beq
\delta \Phi_{WL}(x)\approx \delta \phi(x^\mu+ V^\mu_{WL}(x,x') + V^\mu_C(x')) + \partial_\mu\phi(x^\mu+ V^\mu_{WL}(x,x') + V^\mu_C(x'))(\delta V^\mu_{WL} + \delta V^\mu_C)=0\ ,
\eeq
as in \eqref{Phitrans}.

These expressions can be derived directly from a geometrical construction, while simultaneously generalizing to non-zero initial velocity.  Specifically, we first use the Coulomb dressing to establish a frame at the point $x'=(t',\vec x')$ on the surface defined to linear order by 
\beq\label{Cnorm}
x^\mu=\breve x^\mu + V_C^\mu(\breve x^\mu)
\eeq
with constant $\breve t'$ and varying $\breve x^{i\prime}$.  The spatial vectors of this frame are given by (dropping accents)
\begin{equation} \label{ei}
e^\mu_i = \delta^\mu_i + \partial_i V_C^\mu\ .
\end{equation}
The timelike vector $e_0$ is giving by finding the unit normal to the surface defined by \eqref{Cnorm}; the conditions 
\beq
0 = (\eta_{\mu \nu} + \kappa h_{\mu \nu}) e_0^\mu \partial_i (x^\nu  + V_C^\nu) = \kappa h_{0i} + e_{0i} + \partial_i V_{C0}\quad ,\quad -1 = (\eta_{\mu \nu} + \kappa h_{\mu \nu}) e_0^\mu e_0^\nu
\eeq
give
\begin{equation} \label{e0}
e^\mu_0 = (1 + \frac{\kappa}{2} h_{00}) \hat t^\mu + \delta^{\mu i} (-\kappa h_{0i} - \partial_i V_{C0})\ .
\end{equation}
A geodesic is determined by shooting it from $x'=(t',\vec x')$ with an initial four-velocity specified with respect to this frame.  The choice in \eqref{WLV} corresponds to an initial four-velocity purely in the normal direction, $\partial_0 V^\mu_{WL} = e_0^\mu$.  More generally, we can take initial velocity
\beq
\partial_\tau V^\mu(x,x')_{\Big \vert_{x=x'}} = u^\nu e^\mu_\nu\ .
\eeq

Solving the geodesic equation 
\begin{equation}
\partial_\tau^2 w^\mu + \Gamma^\mu_{uu}=0\,
\end{equation}
for the perturbation $w^\mu$ from the straight-line trajectory, with these initial conditions, gives 
the curve $x^\mu = \breve x^\mu + V^\mu_{WL}(x,x') + V^\mu_C(x')$.  This determines  the worldline dressing
\begin{equation} \label{VWLu}
V_{WL}^\mu(x,x') = \int_0^\tau d \lambda \left[ (\lambda - \tau) \Gamma^\mu_{uu}(x' + \lambda u) + u^\nu e^\mu_\nu(x') \right]\ .
\end{equation}
This can be checked to be gauge invariant, beginning with the gauge transformation for the frame field:
\begin{align}
\delta e^\mu_i &= \partial_i (\delta V_C^\mu) = \kappa \partial_i \xi^\mu, \nonumber \\
\label{framevar}
\delta e^\mu_0 &= -\kappa (\partial_0 \xi_0) \hat t^\mu + \delta^{\mu i} (\kappa \partial_0 \xi_i + \kappa \partial_i \xi_0 - \kappa \partial_i \xi_0) = \kappa \partial_0 \xi^\mu\ .
\end{align}
Then, \eqref{VWLu} can be varied; a $u$-dependent term from varying the Christoffel symbol is cancelled by the second term, using \eqref{framevar}, and we again find the necessary variation,
\eqref{VWLvar}.

In summary the combined Coulomb and worldline dressings can be understood by the following procedure. First we establish an equal-time surface at time $t'$ by the geodesic averaging/Coulomb construction, then to locate a point at a later time $t$ we shoot a geodesic forward  from this surface  for a given proper time in the direction $u^\mu$ specified with respect to the frame at the surface.
One may alternately add a worldline dressing to the Wilson line-dressed operator, analogous to \eqref{WLdressg}.  

The leading-order dressing field resulting from the worldline construction can also be worked out, in analogy to the calculations of \ref{Dressfield}.  We leave the description of this for future work.

\subsection{Gravitational dressing equation}

In Ref.~\cite{Bagan1999a}, dressed field operators for QED were derived from the requirement that in the infinite-mass limit, the equation of motion for the dressed field $\Phi$ should reduce to a first-order equation $i\frac{d}{dt} \Phi =m \Phi$.  This first-order {\it dressing equation} can be shown from the equations of motion to be solved by the worldline dressing \eqref{WLdress}, up to terms of order $1/m$ -- see \eqref{dressing-eq-qed}.
The construction depends on a choice of Lorentz frame, so it selects a preferred dressing for each frame, corresponding to different boosted Coulomb fields.

Also for completeness, we here show that a similar equation is satisfied by the gravitational worldline dressing.
If we study the large-$m$ limit, that means we consider solutions with a rapidly oscillating phase $e^{-i m \breve t}$, where $\breve t$ will be the proper time coordinate used in defining the worldline dressing for gravity.  Here we work in the rest frame of the massive particle, corresponding to choosing a geodesic with zero initial velocity in the preceding subsection, although the discussion may be generalized to non-zero velocity.  
The equation of motion for $\phi$ can be rewritten as
\begin{equation} \label{gravity-large-m}
-i \nabla^\mu \breve t \nabla_\mu \phi - \frac{i}{2} (\nabla^2 \breve t) \phi = m \phi - \frac{e^{- i m \breve t}}{2 m} \nabla^2 (e^{i m \breve t} \phi).
\end{equation}
Here we have used the fact that $\breve t$ measures proper time, $(\nabla^\mu \breve t)^2 = -1$.
To rewrite this as a gravitational dressing equation, we need to reexpress it in terms of the dressed field, and to relate the first term to a time derivative $\partial/\partial \breve t$.  
Recall that the worldline dressed operator $\Phi_{WL}(\breve x)$ is defined as the value of the field at a set of coordinates $\breve x$ with an invariant physical meaning.  Like with the gravitational Wilson line,
$\breve t$ is the proper time along a timelike geodesic, and $\breve x$ labels the different geodesics.
Thus
\beq
\Phi_{WL}(\breve x(x)) = \phi(x)\ ;
\eeq
compare \eqref{brevtox} and \eqref{gwilsdef}.  
Using this, we can reexpress the first term in \eqref{gravity-large-m}
\begin{equation}
-i \nabla^\mu \breve t \nabla_\mu \phi = - i (\nabla^\mu \breve t) (\nabla_\mu \breve x^\nu) \frac{\partial}{\partial \breve x^\nu} \Phi_{WL}(\breve x(x)).
\end{equation}
Since the worldline coordinates are Gaussian-normal, $(\nabla^\mu \breve t) (\nabla_\mu \breve x^\nu) = \eta^{0 \nu}$, and the equation of motion \eqref{gravity-large-m}
becomes
\begin{equation} \label{gravity-dressing-eq}
i \frac{\partial}{\partial \breve t} \Phi_{WL} - \frac{i}{2} (\nabla^2 \breve t) \Phi_{WL} = m \Phi_{WL} - \frac{e^{- i m \breve t}}{2 m} \nabla^2 (e^{i m \breve t} \Phi_{WL})\ ,
\end{equation}
where ${\partial}/{\partial \breve t}$ is defined with the coordinates $\breve x^i$ held fixed.  
This is first-order in time, up to $\calo(1/m)$ corrections.  Dropping the $\calo(1/m)$ term gives the gravitational dressing equation, analogous to that discussed for QED in 
\cite{Bagan1999a}.  

\section{Algebraic structure}\label{algstruct}

An important basic question for a theory is the algebraic structure of its algebra of observables.  In local theories without long-range fields, the algebra has a net structure of subalgebras \cite{Haag} that mirrors the decomposition of the underlying manifold into spacetime regions.  Long range gauge fields lead to additional subtleties with such subalgebras.  We next turn to an examination of some basic aspects of such algebraic structure, focussing on the gauge-invariant observables that we have constructed above, beginning first with QED and then turning to gravity.  

\subsection{Algebra for observables in QED}

\subsubsection{Commutators for Faraday dressing}

Let us first consider the scalar field operator dressed by a Faraday line introduced in section \ref{sec:wils}.
If we consider commutators of $\Phi_{W_z}(x)$ with $\Phi_{W_z}(x')$ at equal times $t=t'$, they will vanish because the operators only involve $\phi$ and $A_z$, all of which commute at equal times (see appendix \ref{app:qed} for basic commutators). However, away from equal times there will be a nonzero commutator 
whenever any part of the two electric strings are causally separated.

To show this we consider one point slightly to the future of the other, which is determined by $[\partial_0 \Phi_{W_z}(x), \Phi_{W_z}(x')]$ at equal times.
Using the definition of $\Phi_{W_z}$, \eqref{Wilsz}, and the identity \cite{Bagan1999a}
\begin{equation} \label{ddteO}
\frac{d}{dt} e^O = e^O \left( \dot O + \frac{1}{2} \left[\dot O, O \right] \right)
\end{equation}
whenever $[\dot O,O]$ is a $c$-number, we find that
\begin{eqnarray} \label{dotPhiWz}
\dot \Phi_{W_z}(x) &=&  e^{i q \int_0^\infty ds A_z(x + s \hat z)} \left[ \dot \phi(x) + i q \phi(x) \int_0^\infty ds \dot A_z(x + s \hat z) + \frac{i}{2} q^2 \phi(x)  \delta^{2} (0) \int_0^\infty ds \right]\ .
\end{eqnarray}
The last term arises from the commutator (see eq.~\eqref{Acommutator}) $[\dot A_z(x), A_z(x')] = -i \delta^{3}(\vec x - \vec x')$; it is formally infinite, and can be interpreted as due to the infinite energy of the singular string.  It can be regulated by cutting off the construction at $z=Z$, and by smearing out the singular string over a small region.  However, since it is proportional to $\Phi_{W_z}$, it can be alternately absorbed by a c-number phase rotation of $\Phi_{W_z}$, and thus also does not contribute to the commutators of present interest.

If we now commute \eqref{dotPhiWz} with $\Phi_{W_z}(x')$ for $x\neq x'$, the contribution of the first term vanishes by $[\dot\phi,\phi]=0$, and we are left with
\begin{eqnarray} \label{Wcommutator}
[\dot \Phi_{W_z}(x), \Phi_{W_z}(x')] 
&=&\Phi_{W_z}(x)\cdot iq \int_0^\infty ds [\dot A_z(x + s \hat z),\Phi_{W_z}(x')] \nonumber \\
&=& i q^2 \Phi_{W_z}(x) \Phi_{W_z}(x') \delta^{(2)}(\vec x_\perp - \vec x'_\perp) \int_{\max(z,z')}^\infty\ dz''.
\end{eqnarray}
The last term is infrared divergent and may be again regulated by cutting off the $z$ integration at finite $Z$.  

To understand the appearance of the nontrivial commutator, imagine first acting on the vacuum with the operator $\Phi_{W_z}(x)$.
This will create a charged particle, and a nontrivial configuration of the electromagnetic field, \eqref{EzWz}, \eqref{Aline}.
If we subsequently act with $\Phi_{W_z}$ at a slightly later point $(t + \delta t, \vec x)$, there is a large phase associated with the introduction of the new charged particle into the preexisting nontrivial field.  The divergence arises since this field does not decay as $z\rightarrow \infty$.  While the Faraday dressing provides a nice intuitive picture of the origin of the non-zero commutator, the divergence can be eliminated by working with dressings like that of Dirac, which are better behaved.

\subsubsection{Commutators for Dirac dressing}

Equal-time commutators of Dirac-dressed operators \eqref{Ddress} likewise vanish at equal times, but can be shown not to vanish as one operator is moved into the future by considering the commutator $[\dot \Phi_D(x),\Phi_D(x')]$.  Specifically, the needed time derivative is
\begin{equation}\label{PhiDdot}
\dot \Phi_D(x) = V_D(x) 
\left( \dot \phi(x) + i q \phi(x) \int d^3 x' \frac{(\vec x'-\vec x)^i}{4\pi|\vec x'-\vec x|^3}\dot A_i(t,\vec x') + \frac{i}{2} q^2\phi(x)\int d^3x'{1\over (4\pi)^2}{1\over |\vec x'-\vec x|^4}\right)\ .
\end{equation}
Here, again, the last piece is interpreted in terms of field energy;  here it is infrared finite, but ultraviolet divergent, though does not contribute to the commutator of interest.  The equal-time commutator arises, as with the Faraday dressing, from the second term in \eqref{PhiDdot}, and takes the form
\begin{eqnarray}\label{Farcom}
[\dot \Phi_D(x), \Phi_D(x')] &=&  \Phi_D(x)\cdot iq  \int d^3x^{\prime\prime}  \frac{(\vxpp-\vec x)^i}{4\pi|\vxpp -\vec x|^3}  [\dot A_i(t,\vxpp), \Phi_D(x')] \nonumber \\
&=& i q^2 \Phi_D(x) \Phi_D(x') \int d^3 \xpp \frac{(\vxpp-\vec x)^i}{4\pi|\vxpp-\vec x|^3} \frac{(\vxpp-\vec x')_i}{4\pi|\vxpp-\vec x'|^3} \nonumber \\
&=& i q^2 \Phi_D(x) \Phi_D(x') \int d^3 \xpp  \frac{1}{4 \pi \abs{\vec x - \vxpp}}  \left( -\nabla^2_{x^{\prime\prime}} \frac{1}{4 \pi \abs{ \vec x' -\vxpp }} \right) \nonumber \\
&=& \frac{i q^2 }{4 \pi \abs{\vec x - \vec x'} } \Phi_D(x) \Phi_D(x')\ .
\end{eqnarray}
In contrast to the divergent commutator of the Faraday-dressed operators, the commutator of the Dirac-dressed operators is nonsingular and decays with distance, reflecting the fact that the Coulomb field spreads out with distance.  Moreover, we see that the coefficient entering the commutator is precisely the Coulomb energy between charges at points $\vec x$ and $\vec x'$.

\subsubsection{Relation to calculations with Dirac brackets}\label{Diraccomp}

Note that \cite{Kabat2012} claims that commutators of the field $\phi$, in axial gauge, vanish at spacelike separation, and argue that this means  microcausality is preserved.  This seems in conflict with the results derived above.  Here we will reanalyze the question, in the Dirac bracket formalism used in \cite{Kabat2012}, and show that the commutators in question are indeed nonvanishing.
To do this we review the Dirac bracket formalism\cite{Hanson1976}, and compare it with the construction used in the present  paper.

First, we recall that the dressed operator $\Phi_{W_z}(x)$ coincides with the value of $\phi(x)$ in axial gauge $A_z = 0$. 
So  the commutation relations of $\Phi_{W_z}(x)$ are given by the commutation relations of $\phi(x)$ in axial gauge, which can be evaluated using the Dirac brackets.  We will encounter a possible ambiguity in this gauge choice, since $A_z$ can really only be set to zero by a gauge transformation vanishing at infinity if $\int dz A_z = 0$; correspondingly, the operator $\phi(x)$ could be accompanied by an electric string running off to $z = \infty$, or to $z = -\infty$, or a linear combination of the two, and these are gauge-inequivalent configurations.

To carry out the Dirac quantization, we begin by treating the components of $A_\mu$ as canonical variables, and find their momenta $\pi^\mu$; these are given in \eqref{piA} with $\alpha \to \infty$. 
Only $\pi^i$ are nonzero, and $\pi^0=0$ gives a primary constraint.
In order that this constraint be preserved in time, we find the Gauss' law constraint, $\partial_i \pi^i = -j^0$, where  $j^0$ is the charge density defined in \eqref{j}.
These constraints Poisson-commute with each other and with the Hamiltonian, and give a system of first-class constraints.

In order to fix a gauge and reduce the system to physical degrees of freedom, we add an additional gauge-fixing constraint, $A_z = 0$. This constraint does not commute with Gauss' law, or with the Hamiltonian.
To preserve the constraint in time, we introduce its time derivative $\pi^z  + \partial_z A_0 = 0$ as a further constraint. Note that this constraint does not commute with the primary constraint $\pi^0 = 0$.
We thus have a system of second-class constraints, given by
\begin{equation}
\pi^0 = 0\ , \quad 
\partial_i \pi^i + j^0 = 0\ , \quad 
A_z = 0\ , \quad 
\pi^z + \partial_z A_0 = 0\ .
\end{equation}

Letting $\chi_i$ label the constraints (there are 4 per point) we can consider the matrix $C_{ij} = \{ \chi_i, \chi_j \}$, written in terms of the Poisson brackets  $\{ A_\mu(x), \pi^\nu(x')\} = \delta_\mu^\nu \delta^3(x - x')$. If this matrix is invertible, then we can form its inverse $C^{ij}$, and define the Dirac brackets,
\begin{equation}
\{ A , B \}_{DB} = \{ A , B \} - \{A , \chi_i \} C^{ij} \{ \chi_j, B \}\ .
\end{equation}
Because  the term $C^{ij}$ is obtained by inverting a differential operator the Dirac bracket is nonlocal.

In the case of axial gauge, the nonzero Poisson brackets are
\begin{eqnarray}
\{\pi^0(x), \pi^z(x') + \partial_z A_0(x')\} & = & \partial_z \delta^3(x - x')\ , \\ 
\{\partial_i \pi^i(x) + j^0(x), A_z(x') \} & = & -\partial_z \delta^3(x - x')\ , \\ 
\{A_z(x), \pi^z (x') - \partial_z A_0(x')] &=& \delta^3(x - x')\ .
\end{eqnarray}
The constraint matrix $C$ can be expressed as a $4 \times 4$ matrix of differential operators, which takes the form
\begin{equation}
C = \left[ 
\begin{array}{cccc}
0 & 0 & 0 & \partial_z \\
0 & 0 & -\partial_z & 0 \\
0 & -\partial_z & 0 & I \\
\partial_z & 0 & -I & 0
\end{array}
\right]\ .
\end{equation}
The general antisymmetric inverse of the constraint matrix is given by
\begin{equation}
C^{-1}(x,x') = 
\left[ 
\begin{array}{cccc}
0 & -\tfrac12 \abs{z - z'} - c z' - d z - e & 0 & \tfrac12 \epsilon(z - z') - c \\
\tfrac12 \abs{z - z'} + c z + d z' + e & 0 & -\tfrac12 \epsilon(z - z') + d & 0 \\
0 & -\tfrac12 \epsilon(z - z') - d & 0 & 0 \\
\tfrac12 \epsilon(z - z') + c & 0 & 0 & 0
\end{array}
\right] \delta^2(\vec x_\perp - \vec x'_\perp)\ .
\end{equation}
We have to choose boundary conditions in order to invert $C$, which determines the constants $c$, $d$ and $e$.
The freedom in inverting the matrix of Poisson brackets is due to the ambiguity in transforming to a gauge such that $A_z = 0$.
The Dirac bracket defined in \cite{Hanson1976} corresponds to the choice $c = d = e = 0$.
These different solutions define different dressed scalar operators.
The simplest way to see this is by considering the Dirac bracket of $\phi$ with the $z$-component of the electric field, which measures the electric flux. 

To find the Dirac bracket of $\phi$ with $E^z$, we need the nonzero Poisson brackets of these quantities with the constraints, which are given by:
\begin{align}
\{ \partial_i \pi^i(x) + j^0(x), \phi(x') \} &= i q \delta^3(x - x') \phi(x')\ , \\
\{ E^z(x), A_z(x') \} &= \delta^3(x - x')\ .
\end{align}
The Dirac bracket is then given by
\begin{align}
\{ E^z(x), \phi(x') \}_{DB} &= - \int d^3 x'' d^3 x''' \{ E^z(x), A_z(x'') \} \left[ -\tfrac12 \epsilon(z'' - z''') - d \right] \delta^2( \vec x''_\perp - \vec x'''_\perp) \{ \partial_i \pi^i(x''') + j^0(x'''), \phi(x') \} \nonumber \\
&= i q \left[ \tfrac12 \epsilon(z - z') + d \right] \delta^2(\vec x_\perp - \vec x'_\perp) \phi(x')\ .
\end{align}
Thus we see for $d = 1/2$ we have a Faraday line of flux $-q$ pointing in the $z$ direction, which is the dressing $\Phi_{W_z}$ of \eqref{EzWz} with the identification $[\ ,\ ] = i \{\ ,\ \}_{DB}$.
The choice $c = 0$, used in \cite{Hanson1976}, gives a particle dressed by two Faraday lines of flux $q/2$ pointing in opposite directions.

We can now consider Dirac brackets between the scalar field $\phi$ and its canonical momentum $\pi_\phi$.
The modification entering the Dirac brackets comes because of the commutators of the matter field with Gauss' law is nonzero.
However $C^{ij}$ is zero when $i$ and $j$ both label the Gauss' law constraint, so these Dirac brackets coincide with the canonical Poisson brackets.
This conclusion was reached in \cite{Kabat2012}, and used as part of an argument for commutativity of $\phi$'s at spacelike separation.

However, while  this argument shows that the Poisson brackets between $\phi$ and $\pi_\phi$ are unmodified by the coupling to the gauge field; it does {\it not} show that $\phi$ commutes with $\phi$ at a later time. 
To see this we can consider the Dirac bracket of $\partial_0 \phi = \pi_\phi^* + i q A_0 \phi$ with $\phi$.
This leads to a nontrivial Dirac bracket due the Poisson bracket of $A_0$ with the primary constraint $\pi^0$:
\begin{align}
\{ \dot \phi(x), \phi(x') \}_{DB} &= - \int d^3 x'' d^3 x''' \{ i q A_0(x) \phi(x), \pi^0(x'') \} \left(-\tfrac12 \abs{z'' - z'''} - c z''' - d z'' - e \right) \delta^2(\vec x''_\perp - \vec x'''_\perp) \{j^0(x'''), \phi(x') \}  \nonumber \\
&= -q^2 \left(\tfrac12 \abs{z - z'} + c z' + d z + e \right) \phi(x) \phi(x')\ .
\end{align}
This indeed agrees with the commutator of the Wilson line dressing \eqref{Wcommutator}, if we take $c = d = \frac12$ and $e = -Z$.

Thus, in summary, while one does find that the equal-time Poisson bracket $\{\phi(x),\pi_\phi(x')\}=0$ for $x\neq x'$,  this does {\it not} imply that $\{\phi(x),\phi(x')\}$ vanishes at spacelike separation, so the operators do not commute in general outside the lightcone.

\subsection{Algebra for observables in gravity}

We finally turn to the important question of the algebra of gauge-invariant operators in gravity.  In section \ref{Gravobs} we described the construction, at linear order in $\kappa$, of diffeomorphism-invariant observables corresponding to a matter field and its gravitational dressing.  In parallel with the preceding discussion on QED, we can now likewise investigate the algebra obeyed by these observables.  While these have only been constructed to leading order in $\kappa$, and thus we will only find the leading-order commutators, these have interesting structure, and of course constrain the all-orders commutators since the latter need to match our results when expanded to this leading order.  

The general form of the gauge-invariant observables we have described is
\beq
\Phi(x)=\phi(x)+ V^\mu(x)\partial_\mu \phi(x) + \kappa^2\Phi_{(2)}(x) +\calo(\kappa^3)\ ,
\eeq
where $V^\mu$ is of order $\kappa$ and explicit examples have been given in section \ref{Gravobs}. We will consider the equal-time commutators $[\Phi,\Phi]$ and $[\dot\Phi,\Phi]$ to leading order, which is to order $\kappa^2$.  At this order and  for $x\neq x'$ these commutators take the form
\beq\label{phicom}
[\Phi(x),\Phi(x')]= [V^\mu(x),V^\nu(x')]\partial_\mu\phi(x)\partial_\nu\phi(x')
\eeq 
and
\beq\label{dotcom}
[\dot\Phi(x),\Phi(x')]= [\dot V^\mu(x),V^\nu(x')]\partial_\mu\phi(x)\partial_\nu\phi(x')+ [V^\mu(x),V^\nu(x')]\partial_\mu\dot\phi(x)\partial_\nu\phi(x')\ .
\eeq

Note that validity of \eqref{phicom} and \eqref{dotcom} requires elimination of other terms that could potentially contribute at order $\kappa^2$.  
First, there is a possible contribution to the commutator \eqref{phicom} of the form 
\begin{equation}
[V^\mu(x), \phi(x')] \partial_\mu \phi(x)
\end{equation}
and similarly with $x \leftrightarrow x'$.
However, $V^\mu(x)$ as we have constructed it depends only on the metric and its first derivative on the constant time slice, so this commutator vanishes.

There is also the possibility of a term in \eqref{dotcom} of the form 
\begin{equation} \label{dotVphi}
[\dot V^\mu(x), \phi(x')] \partial_\mu \phi(x),
\end{equation}
and likewise with $x \leftrightarrow x'$.
Indeed, since $V^0$ contains time derivatives of the spatial metric (see \eqref{WilsV} and \eqref{GC0}), $\dot V^{0}$ contains second time derivatives of $h_{ij}$.
In order to find these commutators at equal time, we have to use the equation of motion  for $h_{\mu \nu}$, \eqref{lingrav}.
It is convenient to choose Feynman gauge and rewrite the equation of motion by subtracting a multiple of the trace as ({\it c.f.} \eqref{lgF})
\begin{equation}\label{meteom}
\ddot h_{\mu \nu} = \vec \nabla^2 h_{\mu \nu} +\frac{ \kappa}{2} \hat T_{\mu \nu}\ , 
\end{equation}
where $\hat T_{\mu \nu}$ is the ``inverse trace-reversed" stress tensor, defined as in \eqref{hatdef}.
This suggests that the nontrivial commutator between $\hat T_{\mu \nu}$ with $\phi$  could potentially lead to a contribution at order $\kappa^2$.
However, the spatial components of $\hat T_{\mu \nu} = \partial_\mu \phi \partial_\nu \phi + m^2 \phi^2 \eta_{\mu \nu}/(D-2)$, which enter $\dot V^0$, do not contain time derivatives of $\phi$, so there is no such contribution to the equal-time commutator between $\dot \Phi$ and $\Phi$; the term in \eqref{dotVphi} vanishes.

Finally, we should consider a possible term in \eqref{phicom} of the form
\begin{equation} \label{phiphi2}
\kappa^2 [\phi(x), \Phi_{(2)}(x')].
\end{equation}
with similar terms appearing in \eqref{dotcom}.
It appears that this term could contain an $\calo(\kappa^2)$ contribution.
However we have defined $\Phi$ as the value of $\phi$ at a point determined in terms of the geometry, so the second-order piece takes the form
\begin{equation}
\kappa^2\Phi_{(2)}(x) = {\frac{1}{2}} V^\mu(x) V^\nu(x) \partial_\mu \partial_\nu\phi(x)+ V^\mu_{(2)} (x) \partial_\mu \phi(x),
\end{equation}
where $V^\mu_{(2)}$ is a nonlocal functional of the metric $h_{\mu \nu}$ of order $\kappa^2$.
Then, the commutator $[\phi(x), h_{\mu \nu}(x')]$ is $\calo(\kappa)$ since it vanishes in the absence of gravitational interactions, so the term in \eqref{phiphi2} is $\calo(\kappa^3)$.

We therefore conclude that \eqref{phicom} and \eqref{dotcom} contain all the terms that can enter the commutators at order $\kappa^2$.

As we have just noted, $V^0$ contains a time derivative of the spatial metric, and this implies a new feature of the algebra, 
 as compared to the gauge-theory case:   there can be a non-zero contribution to the equal-time commutators $[\Phi,\Phi]$.  We will consider the different commutators in turn.

\subsubsection{$[\dot \Phi,\Phi]$}

In this paper, rather than working out the full structure of the commutators in detail, our main focus will be the question of when the commutators are nonzero, and when they become significant.  We first consider commutators $[\dot \Phi,\Phi]$.  These receive contributions in particular when we consider the large-mass limit, where the operators can create massive particles at rest.  

A simplest example of such commutators arises in the Wilson line case, where we find from \eqref{WilsV}, working in $D=4$,
\begin{eqnarray}
[ \dot V_{W_z,z}(x), V_{W_z,z}(x')] &=& \frac{\kappa^2}{4} \int_0^\infty ds \int_0^\infty ds'\; [\dot h_{zz}(x + s \hat z), h_{zz}(x' + s' \hat z)] \nonumber \\
&=& - i \frac{\kappa^2}{8} \delta^2(x_\perp - x_\perp') \int_0^\infty ds\ ,
\end{eqnarray}
using the equal-time commutator (specialized to $D=4$; see \eqref{hcommutator-equaltime}):
\begin{equation}
[\dot h^{\mu \nu}(x), h_{\lambda \sigma}(x')] = -i \left[\delta^\mu_{(\lambda} \delta^\nu_{\sigma) } - \tfrac12 \eta^{\mu \nu} \eta_{\lambda \sigma}\right] \delta^{3}(\vec x - \vec x')\ ,\qquad (t = t', D = 4).
\end{equation}
Thus, from \eqref{dotcom}, there is a nonvanishing contribution to the equal-time commutator
\beq
[\dot\Phi_{W_z}(x),\Phi_{W_z}(x')] = - i \frac{\kappa^2}{8}\partial_z\phi(x)\partial_z\phi(x') \delta^2(x_\perp - x_\perp') \int_0^\infty ds+\cdots\ 
\eeq
where we omit terms proportional to other derivatives of $\phi$.
The divergence here can be regulated by basing the Wilson-line construction on a platform at a large, finite $z=Z$, cutting off the integral at $Z$.  

We next consider commutators proportional to $\dot\phi(x)\dot\phi(x')$; these are the terms that make the leading contribution in the large-mass, zero-momentum limit.  These commutators arise from $[\dot V^0,V^0]$.

An interesting case is that of the commutator of two operators with the Coulomb dressing.  
The time derivative of $V^0_C$ is, using \eqref{meteom} and neglecting the stress tensor terms which we have argued do not contribute to our calculation, and with $\vec r= \vec y-\vec x$,
\begin{eqnarray}\label{vcdot}
\dot V_C^0(x) &=& 
- \frac{\kappa}{4 \pi} \int d^3y \left( \frac{\ddot h_{r r}(y)}{2 r} + \frac{\dot h_{0 r}(y)}{r^2} \right)  \nonumber \\
&=& 
-\frac{\kappa}{4\pi} \int d^3y \left[ h_{\mu \nu} \vec \nabla^2 \left( \frac{\hat r^\mu \hat r^\nu}{2 r} \right) + \frac{\dot h_{0 r}(y)}{r^2} \right] 
- \frac{\kappa}{4 \pi} \oint r^2 d^2 \Omega \left[ \frac{\nabla_r h_{rr}}{2 r} - h_{\mu \nu} \nabla_r \left( \frac{\hat r^\mu \hat r^\nu}{2 r} \right) \right]+ T\ {\rm terms}  \nonumber \\
&=& -\frac{\kappa}{4\pi} \int d^3y \left[ \frac{h^i{}_i - 3 h_{rr}}{r^3} - \frac{2 \pi}{3} h^i{}_i \delta^3(\vec r) + \frac{\dot h_{0r}}{r^2}\right] 
- \frac{\kappa}{4 \pi} \oint r^2 d^2 \Omega \left[ \frac{\nabla_r h_{rr}}{2 r} +  \frac{h_{rr}}{2 r^2} \right]+ T\ {\rm terms} \ . \label{VC0dot}
\end{eqnarray}
Here we have also used the identities \eqref{C4ident}, \eqref{C6ident}.
When evaluating the commutator with $V_C^0$, we can neglect the boundary terms.
This is because the commutator of $h_{\mu \nu}(y)$ with $V_0(x')$ decays as $1/r$, so when we commute $V_0(x')$ with the boundary term in \eqref{VC0dot}, the resulting integrand is $1/r^3$ and hence its integral vanishes as the surface is taken to infinity.

The commutator $[\dot V_C^0(x), V_C^0(x')]$ is the sum of three terms, which we denote by $[\ ]_{1,2,3}$.  
The bulk term from the commutator of $\dot h_{0r}$ with $h_{0r}$ is, with $\vec r'= \vec y' - \vec x'$, 
\begin{eqnarray}
[\ ]_1&=& \frac{\kappa^2}{16 \pi^2} \int d^3 y \int d^3 y' \frac{\hat r^\mu}{r^2} \frac{\hat r'^\nu}{r'^2} \frac{i}{2} \delta_{\mu \nu} \delta^3(\vec y - \vec y') \nonumber \\
&=& \frac{i \kappa^2}{32 \pi^2} \int d^3 y \partial_\mu \left( \frac{1}{r} \right) \; \partial^\mu \left( \frac{1}{r'} \right) \nonumber \\
&=& \frac{i \kappa^2}{8 \pi \abs{x - x'}}\ . \label{VC0dotVC01}
\end{eqnarray}
The piece of the commutator coming from the $1/r^3$ term in \eqref{vcdot} is
\begin{equation}
[\ ]_2
= \frac{i \kappa^2}{32 \pi^2} \int d^3 y \; \frac{1 - 3 (\hat r \cdot \hat r')^2}{r^3 r'}\ .
\end{equation}
This integral is evaluated in appendix \ref{formulas}, and gives
\beq
[\ ]_2=-\frac{i \kappa^2}{6 \pi \abs{x - x'}}\ .
\eeq
Finally, the $\delta$-function term in \eqref{VC0dot} leads to
\begin{eqnarray} \label{VC0dotVC03}
[\ ]_3= \frac{i \kappa^2}{96 \pi \abs{x - x'}}\ .
\end{eqnarray}

Adding these terms gives 
\begin{equation} \label{V0comm}
[\dot V_C^0(x), V_C^0(x')] = [\ ]_1 +[\ ]_2+[\ ]_3= -\frac{i \kappa^2}{32 \pi \abs{x - x'}}\ .
\end{equation}
Then,  if we consider $\Phi_C(x')$ to create a static source, for which $\dot \phi(x') =  i m \phi(x')$, and use \eqref{dotcom}, this leads to a nonzero commutator
\begin{equation}\label{phidotphi}
[\dot \Phi_C(x), \Phi_C(x')] = [\dot V_C^0(x), V_C^0(x')] \dot \phi(x) \dot \phi(x') =  \frac{G m}{\abs{x - x'}} \dot\phi(x) \phi(x')\ .
\end{equation}
Note the comparison between this result and the QED result \eqref{Farcom}; this is even clearer in the static limit $\dot \phi(x) =  i m \phi(x)$.  In higher dimensions, we expect a denominator $\abs{x - x'}^{D-3}$.  Thus in gravity we find nonlocal commutators, and in particular commutators proportional to the gravitational potential.  

\subsubsection{$[\Phi,\Phi]$}

As noted above, a difference between gravity and QED is a nonzero contribution to the equal-time commutators $[\Phi,\Phi]$.  Such contributions arise for both the Wilson line-dressed operators, and the Coulomb-dressed operators.  

These are most easily elucidated by considering the commutator $[\Phi_{W_z}(x),\Phi_C(0)]$.  For simplicity, consider the case where $\vec x=(0,0,z)$.  Then, we see from \eqref{phicom} that these commutators have a contribution from
\begin{eqnarray}\label{WCcomm}
[V^z_{W_z}(x),V_C^0(0)]&=&-\frac{\kappa^2}{8\pi} \int_z^\infty dz'\int d^3y [h_{zz}(0,z'),\dot h_{ij}(y)] \frac{y^iy^j}{2y^3} \nonumber \\
&=&-\frac{i\kappa^2}{32\pi}\int_z^\infty \frac{dz'}{z'} \nonumber \\
&=&-\frac{i\kappa^2}{32\pi}\ln(Z/z)
\end{eqnarray}
where in the last line we have introduced a large-z cutoff $Z$.  
Thus we find
\beq\label{PhiWPhiC}
[\Phi_{W_z}(x),\Phi_C(0)]=-\frac{i\kappa^2}{32\pi}\ln(Z/z)\partial_z\phi(x)\dot\phi(0) +\cdots\ 
\eeq
where we don't include terms proportional to other operators.

While initially surprising, the log has the following explanation.  The time derivative in $V^0$, \eqref{GC0}, creates the perturbation of the spatial metric which we have shown corresponds to linearized Schwarzschild,
\beq
\kappa h_{rr} = \frac{\kappa^2 m}{16\pi r}\ .
\eeq
After acting with the operator $\Phi_C(0)$, the proper distance from a point near infinity is thus corrected by a term logarithmic in $r$; since the Wilson-line construction uses the proper distance, this leads to the log in \eqref{WCcomm}.  Note that this logarithmic term is closely similar to another logarithmic dependence seen in a related physical effect:  the Shapiro time delay.  In higher dimensions, the metric perturbation created by $V_C^0$ is
\beq
\kappa h_{rr}\propto \frac{G_D m}{r^{D-3}}\ ,
\eeq
and so the corresponding correction does not require a cutoff, and varies $\propto z^{4-D}$.  

One can likewise examine the commutator $[\Phi_{W_z}(x),\Phi_{W_z}(x')]$; here one finds a result that is quadratically divergent in the cutoff $Z$ for $D=4$.  

Finally, we consider the commutator of two of the more ``physical" $\Phi_C$'s.  
The commutator of the dressings is 
\begin{eqnarray}
[V_C^0(x), V_C^i(x')] &=&
-\frac{3\kappa^2}{32\pi^2}\int d^3y\int d^3y' \frac{ \hat r^j \hat r^k}{2 r} [\dot h_{jk}(y),h_{lm}(y')]\frac{\hat r'^i \hat r'^l \hat r'^m}{r'^2} \nonumber \\
&=& \frac{3i\kappa^2}{64 \pi^2} \int d^3  y \, \frac{(\hat r \cdot \hat r')^2 - 1/2}{r r'^2} \hat r'^i\ , \label{secondline}
\end{eqnarray}
where $\vec r=\vec y-\vec x$ and $\vec r'=\vec y'-\vec x'$;
in the second line, $\vec y' = \vec y$.
The necessary integral is done in appendix \ref{formulas}, yielding
\beq
[V_C^0(x), V_C^i(x')] =\frac{i\kappa^2}{64 \pi}\frac{x'^i- x^i}{\abs{x-x'}}\ .
\eeq
This result, then, gives a commutator
\beq\label{oddcomm}
[\Phi_C(x),\Phi_C(x')] = -\frac{i\kappa^2}{64\pi}\left[\dot\phi(x)\partial_i\phi(x')+\partial_i\phi(x)\dot\phi(x')\right] \frac{x^i-x'^i}{\abs{x-x'}}\ .
\eeq
This term does not decay with distance, but does vanish with the momentum of the fields, and so vanishes in the static limit.  In higher dimensions, we expect a falloff $\sim |x-x'|^{4-D}$.

Note that the commutator of the form \eqref{oddcomm} can be eliminated by adding a linear combination of $\Delta V^0$ and $\Delta V^i$ appearing in equations \eqref{deltav0}, \eqref{deltavi} to $V^\mu_C$.  For example, the combination
\beq
V^0_N(x) = V^0_C(x)+ \frac{1}{4} \Delta V^0(x)
\eeq
commutes with $V^k(x')$, and so the resulting dressing does not produce a commutator \eqref{oddcomm}.  More generally, one expects a one parameter family of dressings with 
 linear combinations of $\Delta V^0$ and $\Delta V^i$ added to $V_C^\mu$ with the same property.  Note that these will also change the commutators $[\dot V^0, V^0]$, but the latter can be shown to still take the same form as in \eqref{V0comm}, with different numerical coefficients. 

\subsection{Further comments}

We have only evaluated some indicative commutators, which reveal nonlocal behavior, as compared to LQFT, of our gauge-invariant, gravitationally-dressed operators.  Other commutators can be likewise evaluated, with more effort, using similar techniques.  In particular, one can evaluate the commutators of the operators with the worldline dressing, corresponding to more general states of motion of a particle, and find similar results to the simple cases we have shown.

We also note that there are contrary claims in the literature\cite{Kabat2013}, that gravitational dressing does not modify the local properties of commutators of field operators.  The same claim was made\cite{Kabat2012} for QED, for similar reasons.  But, as we have detailed above, a closer inspection of the Dirac brackets for QED fixed to axial gauge explicitly shows noncommutativity that matches that of the dressed operators, giving one confidence in our methods and results.  Thus, we likewise expect that the nonzero commutators for gravity -- which are similar in structure to those of QED -- are also present, and could likewise be derived through a Dirac bracket analysis.

\section{Discussion}

The diffeomorphism-invariant observables that we have constructed in this paper potentially play multiple important roles in better understanding aspects of quantum gravity.  

A first role is to control infrared divergences in scattering.  While that has not been a primary focus of this paper, we have noted that the worldline-dressed operators described in section \ref{WLQED} have been argued by \cite{Bagan1999a,Bagan1999b} to regulate IR divergences in scattering in QED.\footnote{Connections to and issues with the related analysis of \cite{FaKu} are discussed in \cite{HLM}.}  Section \ref{GravWL} has constructed analogous operators in gravity.  Thus, we expect the corresponding gravitational dressings to analogously address IR divergences in gravitational scattering.  We leave development of such a treatment of scattering to future work; for related work see \cite{Akhoury}.

Another very important role is in capturing features of the fundamental structure of quantum gravity.  As we have noted, locality in LQFT is most clearly described in terms of commutativity of subalgebras of gauge-invariant observables associated with spacelike-separated regions of spacetime.  A key question for quantum gravity, if it is a quantum-mechanical theory, is thus what algebraic structure governs its observables, and how this structure relates to possible localization, and reduces to the locality structure of LQFT in the limit $G\rightarrow0$\cite{SGalg}.

While the strong gravitational regime still poses many puzzles, we have found that even the weak-field regime allows us to infer apparently important aspects of this algebraic structure, since non-trivial results can be found perturbatively in $G$, and since any more complete algebra relevant to arbitrarily strong fields must have these contributions at leading-order in an expansion in $G$.  In particular, we have found, at order $G$ in an expansion of commutators of the gauge-invariant observables, obstructions to commutativity of operators associated to different ``regions" of spacetime.  If locality is defined in terms of such commutativity, it fails for the gravitationally-dressed operators we have considered, and in a way that suggests that it will not necessarily be easy to restore with definitions of different operators.  This appears to be an important structural aspect of a quantum theory of gravity, if it is to agree with quantized general relativity in the weak-field regime.

One key piece of information is that of {\it when} the noncommutativity becomes significant; that is expected to be an important characteristic of the ``correspondence boundary" where quantum gravity reduces to LQFT\cite{SG2006,SGalg}.  For example, for two particles of energy $E$, \eqref{phidotphi} indicates that the dressing only makes a small correction to $\dot \Phi(x) \Phi(x')$ when $GE\lesssim \abs{x-x'}$, or, for general $D$, for
\beq\label{locbd}
G_DE\lesssim \abs{x-x'}^{D-3}\ .
\eeq
This is in accord with the {\it locality bound} proposed in \cite{GiLia,GiLib,LQGST}, which stated that LQFT ceases to give an accurate description of the state once the bound \eqref{locbd} is violated.  Note that we have also found significant corrections in the $[\Phi(x),\Phi(x')]$ commutator, which appear to become relevant in a regime where $GpE\gtrsim\abs{x-x'}^{D-4}$, where $p$ is a characteristic momentum of the particles being created.  While interesting, and clearly related to the physical effect described in \eqref{PhiWPhiC}, these commutators can be removed by modifying the dressing, as described in the preceding section.

Notice that we can, at the order $G$ to which we work, achieve commutativity of the gravitational Wilson line operators of \eqref{gwilsdefl} associated to spacelike separated points $x$ and $x'$, {\it if} we run their associate Wilson lines in different directions so that they also stay spacelike separated.  However, such operators do not have a clear identification with a compact spacetime region, and appear to be more clearly associated with a noncompact neighborhood, extending to infinity and containing the flux line.  Moreover, we expect important corrections\cite{SGalg} at higher order in $G$.  First, the infinitesimally-thin Wilson lines have infinite energy density, and thus are expected to receive significant corrections once self-coupling of gravity is taken into account.  This is expected to ``thicken" these Wilson lines.  Moreover, it would appear that the thickness of the corresponding region would grow as the mass or energy sourcing the flux lines increases.  This suggests an important lesson for any putative subalgebra structure:  a monomial in the operator $\Phi_{W_z}(x)$ would appear to be associated with a larger and larger region as the order of the monomial grows; moreover, this also strongly suggests that such a monomial doesn't commute with $\Phi_{W_z}(x')$ for high enough order, spoiling any commutative subalgebra structure.  

While we have treated the case of asymptotically Minkowski space, many of the features we have described should carry over to the case of anti de Sitter space with relatively minor modification.  In particular, one can likewise construct Wilson line operators, associated with ``Fefferman-Graham" gauge, there; for related discussion see \cite{Heemskerk2012,Kabat2013,Almheiri2014}.  These can alternately be averaged over directions, as we have done in section \ref{GravCoul}, to produce a Coulomb-like dressing.  One would then find similar commutator structure for the corresponding operators.  In particular, two Wilson line operators with overlapping Wilson lines are not expected to commute, as in \eqref{Wcommutator}. This differs from  a claim of \cite{Kabat2013}, though we have understood the origin of the conflict in the simple example of QED in section \ref{Diraccomp}.  Also note that for many purposes working in AdS will provide an effective infrared regulator, with characteristic length $\sim R_{\rm AdS}$, to calculations with the flat-space operators described in this paper.

The present discussion also has interesting relations with the similar problem of observables in de Sitter space -- though here nontrivial features are encountered.  In particular, \cite{Giddings2005,GiSl} discussed formulation of observables where field operators are separated by a given geodesic distance; as in the present work, these will create particles together with a gravitational flux line connecting them.  Thus, for far separated particles, or if one particle is taken to be massive and provide a ``platform" with respect to which we measure, the construction for the remaining particle is very similar to the ones we have described.  However, notice that in such a picture field lines appear to terminate on the pair of particles; none can reach asymptotic infinity since space is compact.  This new feature arises from the nontrivial nature of the de Sitter background.  

Past examples\cite{Giddings2005} of diffeomorphism-invariant, approximately local observables such as those just mentioned are {\it relational}, in that the position at which we create or measure a particle is defined in relation to other particles or features of the background state.  Interestingly, the observables constructed in this paper do not require any such local structure to define location; the location of the field operator is defined in relation to structure at infinity.  Thus these are still relational, though in a somewhat different fashion.

A final role to consider for such observables is their connection to observation or experiment.  The observables of this paper, like those of \cite{Giddings2005,GiSl} are observables in the usual mathematical sense of quantum mechanics -- they are Hermitian, gauge-invariant operators on the Hilbert space.   However, they do not have an a priori connection to observations made by ``observers inside the system" and thus were referred to as {\it q-observables} in \cite{GiSl2}.  Some such q-observables are expected to be related to observations such observers can make (thus to what experimental physicists would call ``observables"); further development of this story is left for future work.

\vskip.1in
\noindent{\bf Acknowledgements.} 
We wish to thank D. Harlow, J. Hartle,  D. Kabat, and D. Marolf for discussions. The work of SBG was supported in part by the Department of Energy under Contract DE-SC0011702, by  grant FQXi-RFP3-1330 from the Foundational Questions Institute (FQXi)/Silicon Valley Community Foundation, and by  the National Science Foundation under Grant No. NSF PHY11-25915 to the Kavli Institute of Theoretical Physics, whose hospitality during the workshop ``Quantum gravity foundations: UV to IR"  is also gratefully acknowledged.

\appendix

\section{QED basics} \label{app:qed}

Here we collect some basic formulas relevant to quantization of QED. 

The Lagrangian of QED takes the form
\begin{equation}
\call_{\rm QED} = -\frac14 F^{\mu \nu} F_{\mu \nu} - \frac{1}{2\alpha} (\partial_\mu A^\mu)^2 + \call_m\ .
\end{equation}
The second term is a ``gauge-fixing" (really, gauge-invariance breaking) term; gauge transformations act as 
\beq
A_\mu(x)\rightarrow A_\mu(x)-\partial_\mu\Lambda(x)\ .
\eeq
Gauge symmetry is restored for $\alpha=\infty$, $\alpha=1$ gives ``Feynman gauge," and $\alpha\rightarrow0$ gives ``Lorenz" or ``Landau gauge." The third term is the matter Lagrangian.  The corresponding equations of motion are
\begin{equation} \label{eom}
\partial_\nu F^{\mu \nu} - \frac{1}{\alpha} \partial^\mu \partial_\nu A^\nu = -\square A^\mu + \left( 1-\frac{1}{\alpha} \right) \partial^\mu \partial_\nu A^\nu =  j^\mu.
\end{equation}
where $j^\mu = \frac{\delta}{\delta A_\mu}\int d^4 x \call_m$ is the current; here $\square=\partial_\mu\partial^\mu$.  

A particular matter Lagrangian is that for a charged scalar,
\begin{equation}
\call_m = -\abs{D_\mu \phi}^2 - m^2 \abs{\phi}^2,
\end{equation}
with gauge transformation
\begin{equation}
\phi(x) \to e^{-i q \Lambda(x)} \phi(x).
\end{equation}
and with covariant derivative
\beq
D_\mu \phi = \partial_\mu \phi - i q A_\mu \phi\ .
\eeq
The corresponding current is
\begin{equation} \label{j}
j^\mu = -i q [\phi^* D^\mu \phi - (D^\mu \phi)^* \phi]\ .
\end{equation}

The canonical momenta are
\beq \label{phi-canonical-momenta}
\pi_\phi= (D_0 \phi )^*\quad ,\quad \pi_\phi^* = D_0\phi
\eeq
and
\begin{equation} \label{piA}
\pi^i = -\partial^0 A^i + \partial^i A^0= -F^{0i}=-E^i, \qquad \pi^0 = \frac{1}{\alpha} \partial_\mu A^\mu\ .
\end{equation}
At $\alpha=\infty$, $\pi^0$ of course vanishes, yielding a constraint. 
The equal-time commutators are
\beq
[\pi_\phi(x),\phi(x')]_{\big|_ {t=t'}}=-i\delta^{{D-1}}({\vec x} -{\vec x}')=[\pi_\phi^*(x),\phi^*(x')]_{\big|_ {t=t'}}
\eeq
and
\begin{equation}\label{EMet}
[\pi^\mu(x), A_\nu(x')]_{\big|_ {t=t'}} = -i \delta^\mu_\nu \delta^{{D-1}}(\vec x - \vec x'). 
\end{equation}

These commutation relations provide initial data for the unequal-time commutators.  In the free limit, these satisfy the free equations of motion.  For the free scalar, we have
\begin{equation}
[\phi(x), \phi^*(x')] =  i \Delta(x-x').
\end{equation}
where $\Delta$ is the massive Pauli-Jordan function, satisfying 
\begin{equation}
(\square - m^2)\Delta(x) = 0\ , \qquad 
\Delta(x)_{\big|_{t = 0}} = 0\ ,\qquad 
\partial_t \Delta(x)_{\big|_ {t=0}}  = -\delta^{3}(\vec x)\ .
\end{equation}
Note also that
\beq
\Delta(x-x') = G_a(x,x') - G_r(x,x')\ ,
\eeq
where $G_r$ and $G_a$ are the retarded and advanced Green functions, respectively,
\beq
G_r(x,x')=i\theta(t-t')\langle 0|[\phi(x), \phi^*(x')]|0\rangle\quad,\quad G_a(x,x')=-i\theta(t'-t)\langle 0|[\phi(x), \phi^*(x')]|0\rangle\ .
\eeq

The commutators for the electromagnetic field may be written in terms of the massless Pauli-Jordan function (again, advanced minus retarded Green function), satisfying
\begin{align}
\square D(x) &= 0\ ,  \\
D(-x) &= -D(x)\ ,  \\
D(x)_{\big|_ {t=0}} &= 0\ ,  \\
\partial_t D(x)_{\big|_ {t=0}} &= -\delta^{{D-1}}(\vec x)\ .
\end{align}
In $4$ dimensions, $D(x)$ is given by
\begin{equation} \label{Dscalar}
D(x) = -\frac{1}{2 \pi} \epsilon(t) \delta(x^2),
\end{equation}
where $\epsilon(t)$ is the sign function.
The commutators become
\begin{equation} \label{Acommutator}
[A_\mu(x), A_\nu(x')] =  i D_{\mu \nu}(x-x')
\end{equation}
where 
\begin{equation}\label{EMprop}
D_{\mu \nu}(x) = \eta_{\mu \nu} D (x) + (1 - \alpha) \partial_\mu \partial_\nu E(x),
\end{equation}
and $\square E = -D$.  More explicitly, $E(x)$ is the Green function for the operator $\square^2$ with the 
boundary conditions
\begin{equation}
E(x)_{\big|_ {t=0}} = \partial_t E(x)_{\big|_ {t=0}} = \partial_t^2 E(x)_{\big|_ {t=0}} = 0\ , \qquad \partial_t^3 E(x)_{\big|_ {t=0}} = - \delta^{D-1}(\vec x)\ .
\end{equation}
In $D = 4$, it is constant in the forward lightcone, and in the backward lightcone,  and is given explicitly by:
\begin{equation}
E(x) = -\frac{1}{8\pi} \epsilon(t) \theta(-x^2),
\end{equation}
Eq.~\eqref{EMprop} can be verified by checking that the resulting commutator satisfies the equation of motion,
\begin{equation}
\left[\delta^\mu_\lambda\square  + \left( \frac{1}{\alpha} - 1 \right) \partial^\mu \partial_\lambda\right] D_{\mu \nu}(x) = 0\ ,
\end{equation}
with initial conditions given by:
\begin{equation}
[A_0(x), \dot A_0(x')] = -i \alpha \delta^{{D-1}}(\vec x - \vec x')\ , \qquad
[A_i(x), \dot A_j(x')] = i \delta_{ij} \delta^{{D-1}}(\vec x - \vec x')\ ,
\end{equation}
which follow from \eqref{EMet}.

Note that the quantity $B=\partial_\mu A^\mu$ generates gauge transformations on both the electromagnetic and matter fields:
\begin{equation} \label{B}
i[B(x'),A_\mu(x) ] =  \alpha \partial_\mu D(x'-x) \qquad , \qquad i[B(x'),\phi(x)] = i\alpha q D(x' - x) \phi(x).
\end{equation}
This is an infinitesimal gauge transformation, 
\begin{equation}\label{Btrans}
i[B(x'), A_\mu(x) ] =  \delta_\Lambda A_\mu(x) = - \partial_\mu \Lambda(x)\qquad , \qquad i[B(x'), \phi(x) ] = \delta_\Lambda \phi(x) = -i q \Lambda(x) \phi(x')
\end{equation}
with $\Lambda(x) = \alpha D(x - x')$.  Here the commutators with $A_\mu$ follow from \eqref{Acommutator}.  To check the $\phi$ commutator, 
 we first use the equal-time commutation relations to show that it holds at equal times,
\beq
[B(x'),\phi(x) ]_{\big|_ {t=t'}} = 0\qquad , \qquad [\partial_0 B(x'), \phi(x) ]_{\big|_ {t=t'}} = -\alpha q \delta^3 (\vec x - \vec x') \phi(x)\ ,
\eeq
where we have used the equation of motion to write $\partial_0 B = \alpha (\partial_i \pi^i + j^0)$.
The identity \eqref{B} then follows at unequal times using the fact that $B$ satisfies the free equation of motion $\square B = 0$ when the current $j^\nu$ is conserved.

Thus, a gauge-invariant operator $\Phi$ constructed out of the electromagnetic field and the scalar field $\phi$ will commute with $B(x)$ .
The physical states of the theory are those annihilated by the positive-frequency part $B^+$ of $B$:
\begin{equation}
B^+(x) \ket{\psi} = 0 \qquad \Leftrightarrow \qquad \ket{\psi} \text{ is a physical state}.
\end{equation}
Then if $\Phi$ is a gauge-invariant operator, it will commute with the positive-frequncy part of $B$, and hence maps physical states to physical states.

\section{Gravity basics} \label{app:gravity}

Here we collect some basic formulas relevant to perturbative quantization of gravity.

The scalar Lagrangian density for Einstein gravity takes the form
\begin{equation}\label{gravL}
\call_{\rm grav} = \frac{2}{\kappa^2}  R +\call_{gf}+ \call_m\ ,
\end{equation}
where $\kappa^2=32\pi G_D$, $G_D$ is the $D$-dimensional Newton's constant, $R$ is the scalar curvature and $\call_m$ is the matter Lagrangian; a particular example is that for a scalar with mass $m$,
\beq
\call_m=-\hf  \left[(\nabla\phi)^2+m^2\phi^2\right]\ .
\eeq
The second term in \eqref{gravL}, $\call_{gf}$, denotes a ``gauge-fixing" (really, gauge-invariance breaking) term.  If one picks a background metric $g^0$, one useful choice is
\beq
\sqrt{\abs g} \call_{gf}=-\frac{1}{\alpha\kappa^2}\frac{\sqrt{\abs{g^0}}}{\abs{g}}\left[\nabla^0_\mu\left(\sqrt{|g|} g^{\mu\nu}\right)\right]^2\ ,
\eeq
where $\nabla^0$ denotes the covariant derivative with respect to $g^0$ and $g^0$ is used for the contraction of the $\nu$ index.
Gauge symmetry is restored for $\alpha=\infty$, $\alpha=1$ is ``Feynman gauge," and $\alpha\rightarrow0$ is an analog of ``Landau gauge," which enforces the de Donder gauge condition
\beq
\nabla^0_\mu\left(\sqrt{|g|} g^{\mu\nu}\right) = 0\ .
\eeq
When $g^0$ is the flat metric, this reduces to the usual harmonic gauge condition, which can be expressed in any of the equivalent forms:
\begin{equation}
\partial_\mu \left( \sqrt{|g|} g^{\mu \nu} \right) = 0\ ,
\qquad g^{\mu \nu} \Gamma^\alpha_{\mu \nu} = 0 \ , \qquad \square X^\mu = 0\ .
\end{equation}
The latter says that the coordinates $X^\mu$ are harmonic functions of spacetime.

For the purposes of this paper we primarily focus on the linearization of gravity about flat space.  The metric perturbation is defined by
\beq
g_{\mu\nu}=\eta_{\mu\nu} + \kappa h_{\mu\nu}\ .
\eeq
For the linearized theory we need the quadratic-order expansion of $\sqrt{|g|}\call_{EH}= 2\sqrt{|g|} R/\kappa^2$ in $h$.  Various formulas simplify if we define the ``trace-reversed" metric perturbation,
\beq
{\bar h}_{\mu\nu}= h_{\mu\nu}-\hf \eta_{\mu\nu} h\ ,
\eeq
with $h=\eta^{\mu\nu}h_{\mu\nu}$; the inverse to trace reversal, in $D$ spacetime dimensions, is
\beq\label{hatdef}
\hat{\bar h}_{\mu\nu}= {\bar h}_{\mu\nu} -\frac{1}{D-2} \eta_{\mu\nu} {\bar h} = h_{\mu\nu}\ .
\eeq
The quadratic part of the Einstein-Hilbert action can then be simplified to 
\beq
\left(\sqrt{|g|}\call_{EH}\right)_2=-\hf \partial_\sigma h_{\mu\nu} \partial^\sigma{\bar h}^{\mu\nu} + \partial^\lambda {\bar h}_{\lambda \mu} \partial_\nu {\bar h}^{\nu \mu} +{\rm t.d.}
\eeq 
where indices are raised with the flat metric $\eta$ and  the last term is a total derivative.  Likewise, to quadratic order, the gauge-fixing term gives
\beq
\left(\sqrt{|g|}\call_{gf}\right)_2= -\frac{1}{\alpha} \left(\partial^\lambda {\bar h}_{\lambda \mu}\right)^2\ ,
\eeq
and so the combined quadratic action for gravity is (dropping total derivatives)
\beq
\label{quadact}
\left[\sqrt{|g|}\left(\call_{EH}+\call_{gf}\right)\right]_2 = -\hf \partial_\sigma h_{\mu\nu} \partial^\sigma{\bar h}^{\mu\nu} +\left(1-\frac{1}{\alpha}\right)\left(\partial^\lambda {\bar h}_{\lambda \mu}\right)^2\ .
\eeq

This action exhibits the simplicity of Feynman gauge, $\alpha=1$.   
Here, the canonical conjugate to $h_{\mu \nu}$ is $\dot {\bar h}^{\mu \nu}$, so we have the equal-time commutation relations
\begin{equation} \label{hhbarcommutator-equaltime}
[h_{\mu \nu}(x), \dot{\bar{h}}^{\lambda\sigma}(x')]_{\big|_ {t=t'}} = i \delta_{\mu\nu}^{\lambda\sigma} \delta^{D-1}(\vec x-\vec x')\ ,
\end{equation}
where we define
\beq
\delta_{\mu\nu}^{\lambda\sigma}=\delta_\mu^{(\lambda}\delta_{\nu}^{\sigma)}
\eeq
with symmetrization convention $A^{(\lambda\sigma)} =(A^{\lambda\sigma} +A^{\sigma\lambda} )/2$.
Equivalently without using the trace-reversed field,
\beq \label{hcommutator-equaltime}
[h_{\mu \nu}(x), \dot{{h}}^{\lambda\sigma}(x')]_{\big|_ {t=t'}} = i\left( \delta_{\mu\nu}^{\lambda\sigma}-\frac{\eta_{\mu \nu} \eta^{\lambda \sigma}}{D-2}\right)\delta^{D-1}(\vec x-\vec x')\ .
\eeq

When $\kappa = 0$, the field equation reduces to $\square \bar h_{\mu \nu} = 0$, from which we find the unequal-time Feynman-gauge commutation relations for fields in the interaction picture:
\begin{equation}
[\bar h_{\mu \nu}(x), h^{\lambda\sigma}(x')] =  i \delta_{\mu\nu}^{\lambda\sigma}D(x-x')\ ,
\end{equation}
or
\begin{equation} \label{hcommutator}
[h_{\mu \nu}(x), h^{\lambda\sigma}(x')] = {i} \left(\delta_{\mu\nu}^{\lambda\sigma} - \frac{\eta_{\mu \nu} \eta^{\lambda \sigma}}{D-2} \right) D(x-x')\ ,
\end{equation}
with $D(x)$ given in appendix \ref{app:qed}.
Alternately, the momentum-space two-point function is
\beq\label{gravprops}
\left\langle h_{\mu\nu}(p) {\bar h}^{\lambda\sigma}(p') \right\rangle= -\frac{i}{p^2} \delta_{\mu\nu}^{\lambda\sigma} (2\pi)^D\delta^D(p+p')\ ,
\eeq
with the corresponding correlators for the metric found via the transformation \eqref{hatdef}.

These expressions can be generalized to $\alpha\neq1$; let us introduce the variable
\beq
\beta=1-\frac{1}{\alpha}\ .
\eeq 
Then, the quadratic action \eqref{quadact} takes the form
\beq 
\left[\sqrt{\abs g}\left(\call_{EH}+\call_{gf} \right)\right]_2= \hf h^{\mu\nu} L_{\mu\nu}{}^{\lambda\sigma}(\beta) h_{\lambda\sigma}
\eeq
where $L_{\mu\nu}{}^{\lambda\sigma}(\beta)$ is the second-order linear operator defined by
\beq
L_{\mu\nu}{}^{\lambda\sigma}(\beta) = \left( \delta_{\mu\nu}^{\lambda\sigma} - \frac12 \eta_{\mu \nu} \eta^{\lambda \sigma} \right) \partial^\rho \partial_\rho
- 2 \beta \left(\delta^\alpha_{(\mu}\partial_{\nu)} -\hf \eta_{\mu\nu}\partial^\alpha\right)\left(\delta_\alpha^{(\lambda}\partial^{\sigma)}-\hf\eta^{\lambda\sigma}\partial_\alpha\right)\ .
\eeq 
Defining the stress tensor as
\beq
T_{\mu\nu} = -\frac{2}{\sqrt{\abs g}}\frac{\delta S_m}{\delta g^{\mu\nu}}\ ,
\eeq
the linearized gravitational equations then take the form
\beq\label{lingrav}
L_{\mu\nu}{}^{\lambda\sigma}(\beta)h_{\lambda\sigma} = - \frac{\kappa}{2} T_{\mu\nu}\ ,
\eeq
where in this equation the metric in $T_{\mu\nu}$ gets replaced with $\eta$ to leading order.
Then, the propagator for the metric takes the form
\beq\label{mettp}
\langle T h_{\mu\nu}(x) h^{\lambda\sigma}(x')\rangle = i(L^{-1})_{\mu\nu}{}^{\lambda\sigma}(x,x')\ .
\eeq
Note that the linearized gravitational equations \eqref{lingrav} simplify to
\beq\label{lgF}
\square \bar h_{\mu \nu} = - \frac{\kappa}{2} T_{\mu \nu}
\eeq
in Feynman gauge ($\beta=0$).

One can also work out the canonical momenta and commutators for the metric.  From \eqref{quadact}, one finds
\begin{align} \label{pigrav}
\pi^{0\mu}&=\dot {\bar h}^{0\mu}-\beta\partial_\nu {\bar h}^{\nu\mu}\ ,\nonumber\\
\pi^{ij} &= \dot {\bar h}^{ij} - \beta\partial_\nu\bar h^{\nu 0}  \delta^{ij}\ .
\end{align}
When $\beta = 1$ there is no gauge fixing, so the momenta $\pi^{\mu 0}$ do not contain time derivatives, and give the expected constraints.
For general $\beta$ the canonical commutators are
\beq \label{gravcc}
[\pi^{\mu\nu}(x),h_{\lambda\sigma}(x')]_{\big|_ {t=t'}}=-i\delta^{\mu\nu}_{\lambda\sigma}\delta^{D-1}(\vec x -\vec x')\ .
\eeq 
When $\beta = 0$, this agrees with the Feynman gauge result \eqref{hhbarcommutator-equaltime}.

These commutation relations provide initial data for the unequal-time commutators, which in the free limit satisfy the free equations of motion, \eqref{lingrav} with $T=0$. These are
\beq\label{gravgenc}
[h_{\mu\nu}(x),\bar h^{\lambda\sigma}(x')]=iD_{\mu\nu}{}^{\lambda\sigma}(x-x')
\eeq
with
\beq\label{gravD}
 D_{\mu\nu}{}^{\lambda\sigma}(x)=\delta_{\mu\nu}^{\lambda\sigma}D(x) + 2(1-\alpha) \left[ \delta_{(\mu}^{(\lambda}\partial^{\sigma)}\partial_{\nu)} -\hf \eta^{\lambda\sigma} \partial_\mu\partial_\nu\right]E(x)
 \eeq
 and $D(x)$ and $E(x)$ given in appendix \ref{app:qed}.  This can be checked by verifying that 
 \beq
 L_{\mu\nu}^{(x)}{}^{\lambda\sigma} D_{\lambda\sigma}{}^{\gamma\delta}(x,x')=0
 \eeq
 and that $D_{\mu\nu}{}^{\lambda\sigma}$ satisfies the initial conditions implied by 
 \beq
 [h_{\mu\nu}(x),h_{\lambda\sigma}(x')]_{\big|_ {t=t'}}=0
 \eeq
 and \eqref{gravcc}.

The gauge condition $b^\nu = \partial_\mu \bar h^{\mu \nu}$ generates infinitesimal diffeomorphisms, just as the gauge condition in QED generates infinitesimal gauge transformations ({\it c.f.} \eqref{Btrans}).
From \eqref{gravgenc} and \eqref{gravD} we find the commutator
\begin{equation} \label{Bgrav}
i[b^\sigma(x'),h_{\mu \nu}(x)] = 
-\partial_\mu \xi^{(\sigma)}_\nu(x) - \partial_\nu \xi^{(\sigma)}_\mu(x)\ ,
\end{equation}
with the infinitesimal diffeomorphism generated by the vector fields
\begin{equation}
\xi^{(\sigma)}_\mu(x) = \frac{\alpha}{2} \delta^\sigma_\mu D(x - x').
\end{equation}
One can also see that 
\beq\label{phigrav}
i[b^\sigma(x'), \phi(x)] = - \kappa \xi^{(\sigma)}_\mu (x) \partial^\mu \phi(x)\ .
\eeq
To show this, one first checks the equal-time versions of it, 
\begin{align}
i[b^\sigma(x'), \phi(x)]_{\big|_ {t=t'}}
&= 0 \ , \nonumber \\
i[\dot b^\sigma(x'), \phi(x)]_{\big|_ {t=t'}}
&= {i\frac{\alpha\kappa}{2} [ T^{0 \sigma}(x'),\phi(x)]}
= - \frac{ \alpha\kappa}{2} \partial^\sigma \phi(x) \delta^{D-1}(x - x')\  \label{BgravIC}
\end{align}
where the latter commutator follows from the equations of motion \eqref{lingrav}.  Then, by taking the divergence of the same equations of motion, we find 
\begin{equation}
\square b^\sigma = 0\ 
\end{equation} 
when the stress tensor is conserved.  Thus 
since \eqref{phigrav} satisfies this equation of motion in $x'$ and the initial conditions \eqref{BgravIC} at $t=t'$, it holds for all $x'$.

In particular, this implies that if $\Phi$ is any operator invariant under linearized diffeomorphisms, then the linearized metric $\tilde h$ defined to leading order in $\kappa$ by
\begin{equation}
\tilde h_{\mu \nu}(x)\Phi(x') = [h_{\mu \nu}(x), \Phi(x')]
\end{equation}
satisfies the gauge condition $\partial_\mu \tilde {\bar h}^{\mu \nu} = 0$.

\section{Some useful formulas}\label{formulas}

Here we collect some derivative formulas and integrals used in the main text.
It will be useful to introduce a radial vector field $\hat r^\mu$ such that $\hat r^\mu \partial_\mu = \partial_r$; we will denote the spatial metric by $q_{\alpha\beta}$
Then we have the identities (given for $D=4$, though easily generalized to $D\neq 4$)
\begin{align}
\partial_\mu\partial_\nu r &= \partial_\mu \hat r_\nu = \frac{1}{r} (q_{\mu \nu} - \hat r_\mu \hat r_\nu), \\
\partial_\mu \hat r^\mu &= \frac{2}{r} \\
\square \hat r^\mu 
&= -\frac{2}{r^2} \hat r^\mu \\
\label{C4ident}
\partial_\mu \left( \frac{\hat r_\alpha \hat r_\beta}{r} \right) 
&= \frac{q_{\mu \alpha} \hat r_\beta + q_{\mu \beta} \hat r_\alpha - 3 \hat r_\mu \hat r_\alpha \hat r_\beta }{r^2}  \\
\partial_\mu \left( \frac{\hat r^\mu \hat r_\beta}{r} \right) 
&= \frac{\hat r_\beta}{r^2} \\
\label{C6ident}
\square \left( \frac{\hat r_\alpha \hat r_\beta}{r} \right) 
&= 2 \frac{q_{\alpha \beta} - 3 \hat r_\alpha \hat r_\beta}{r^3} - \frac{4 \pi}{3} q_{\alpha \beta} \delta^3(\vec x) \\
\label{3deriv}
\partial_\alpha \partial_\beta\partial_\gamma r 
&= \frac{3}{r^2} (\hat r_\alpha \hat r_\beta \hat r_\gamma - \hat r_{(\alpha} q_{\beta\gamma)}) \\
\partial_\alpha \partial_\beta \partial_\gamma \frac{1}{r}
&= \frac{9 \hat r_{(\alpha} q_{\beta\gamma)} - 15 \hat r_\alpha \hat r_\beta \hat r_\gamma}{r^4}
 - \frac{12 \pi}{5} q_{(\alpha \beta} \partial_{\gamma)} \delta^3(\vec x)
\ ,
\end{align}
where in the last two equations $\hat r_{(\alpha} q_{\beta\gamma)}= (\hat r_{\alpha} q_{\beta\gamma}+\hat r_{\beta} q_{\gamma\alpha}+\hat r_{\gamma} q_{\alpha\beta})/3$.

Calculating the commutators in section \ref{algstruct} requires evaluation of certain integrals.  The first is 
\beq
\int d^3 y \; \frac{1 - 3 (\hat r \cdot \hat r')^2}{r^3 r'}\ 
\eeq
where $\vec r = \vec y - \vec x$, $\vec r'= \vec y - \vec x'$.
This is evaluated by choosing spherical coordinates based at $\vec x=0$, and with polar direction defined as that of $\vec d=\vec x' -\vec x$.  
In particular, we then find
\begin{equation}\label{rrprime}
(\hat r \cdot \hat r')^2 = 1 - \frac{d^2}{r'^2} (1 - \cos^2\theta ).
\end{equation}
We then expand $1/r'$ and $1/r'^3$ in Legendre and Gegenbauer polynomials respectively:
\begin{eqnarray} \label{Legendre}
\frac{1}{r'} &=& \frac{1}{\sqrt{r^2 - 2 d r
 \cos\theta + d^2}} = \begin{cases} 
\frac{1}{r} \sum_{l =0}^\infty P_l(\cos\theta) \left( \frac{d}{r} \right)^l & d < r \\
\frac{1}{d} \sum_{l=0}^\infty P_l(\cos\theta) \left( \frac{r}{d} \right)^l & r< d
\end{cases} \\
\frac{1}{r^{\prime3}} &=& \frac{1}{(r^2 - 2 d r
 \cos\theta + d^2)^{3/2}} = \begin{cases} 
\frac{1}{r^3} \sum_{l=0}^\infty C^{3/2}_l(\cos\theta) \left( \frac{d}{r} \right)^l & d < r \\
\frac{1}{d^3} \sum_{l=0}^\infty C^{3/2}_l(\cos\theta) \left( \frac{r}{d} \right)^l & r < d \label{Gegenbauer}
\end{cases}
\end{eqnarray}
The angular integral picks out the $l = 0$ term of each sum, and we find
\begin{eqnarray}
\int d^3 y \; \frac{1 - 3 (\hat r \cdot \hat r')^2}{r^3 r'}&=& 2\pi \int r^2 dr d \cos(\theta) \left[\frac{-2}{r^3 r'} + 
\frac{3 d^2}{r^3 r'^3} (1 - \cos\theta^2) \right]\nonumber \\
&=& 2\pi \left[ \int_0^d dr \left( -\frac{4 }{r d} + \frac{4}{r d}\right) + \int_d^\infty dr \left( -\frac{4}{r^2} + \frac{4 d^2}{r^4} \right) \right]\nonumber  \\
&=& -\frac{16\pi }{3\abs{x - x'}}\ . 
\end{eqnarray}

Another needed integral is
\beq
I^i= \int d^3  y \, \frac{a (\hat r \cdot \hat r')^2 + b}{r r'^2} \hat r'^i\ ,
\eeq
where $a$, $b$ are fixed constants.
For this we center spherical coordinates about the point $x'$, with the polar axis determined by $\vec d= \vec x - \vec x'$.  By symmetry, the integral must be proportional to $\hat d^i$, with coefficient $\hat d^i I^i$.  We also use \eqref{rrprime}, with $\vec r \leftrightarrow \vec r'$, giving
\beq
\hat d^i I^i = 2\pi \int dr' d\cos\theta \left[\frac{a + b}{r} - a \frac{d^2(1-\cos^2\theta)}{r^3}\right]\cos\theta\ .
\eeq
As before, we expand $1/r$ and $1/r^3$ in Legendre and Gegenbauer polynomials, using \eqref{Legendre}, \eqref{Gegenbauer} with $r \leftrightarrow r'$.
In both cases the $\theta$ integral picks out the $l=1$ term of the sum:
\begin{equation}
\int_{-1}^1 d \cos\theta \cos\theta P_l(\cos\theta) = \frac{2}{3} \delta_{l1}, \qquad \int_{-1}^1 d \cos\theta \cos\theta C^{3/2}_l(\cos\theta) (1 - \cos\theta^2) = \frac{4}{5} \delta_{l1}.
\end{equation}
Substituting this into the above integral, we find 
\beq
\hat d^i I^i = 2\pi \left[ \int_0^d dr' \left (\frac{2}{3}(a + b) - \frac{4}{5} a \right) \frac{r'}{d^2} + \int_d^\infty dr' \left(\frac{2}{3}(a + b) \frac{d}{r'^2} - \frac{4}{5} a \frac{d^3}{r'^4} \right) \right] = 2 \pi \left( \frac{a}{3} + b \right)
\eeq
and thus
\beq
\int d^3  y \, \frac{a (\hat r \cdot \hat r')^2 + b}{r r'^2} \hat r'^i\ =  2 \pi \left( \frac{a}{3} + b \right) \frac{\vec x^i-\vec x'^i}{\abs{x-x'}}\ .
\eeq

\bibliographystyle{utphys}
\bibliography{QG-obs}

\providecommand{\href}[2]{#2}\begingroup\raggedright\begin{thebibliography}{10}

\bibitem{UQM}
S.~B. Giddings, ``{Universal quantum mechanics},''
  \href{http://dx.doi.org/10.1103/PhysRevD.78.084004}{{\em Phys.Rev.}
  {\bfseries D78} (2008) 084004},
\href{http://arxiv.org/abs/0711.0757}{{\ttfamily arXiv:0711.0757 [quant-ph]}}.

\bibitem{Dewi1}
B.~S. DeWitt, ``{The Quantization of geometry},'' in {\em Gravitation: An
  Introduction to Current Research}, W.~L, ed., pp.~266--381.
\newblock Wiley, New York,
1962.
\newblock

\bibitem{Dewi2}
B.~S. DeWitt, ``{Quantum Theory of Gravity. 1. The Canonical Theory},''
\href{http://dx.doi.org/10.1103/PhysRev.160.1113}{{\em Phys.Rev.} {\bfseries
  160} (1967) 1113--1148}.

\bibitem{TsWo}
N.~Tsamis and R.~Woodard, ``{Physical Green's Functions in Quantum Gravity},''
\href{http://dx.doi.org/10.1016/0003-4916(92)90301-2}{{\em Annals Phys.}
  {\bfseries 215} (1992) 96--155}.

\bibitem{Rove}
C.~Rovelli, ``{Partial observables},''
  \href{http://dx.doi.org/10.1103/PhysRevD.65.124013}{{\em Phys.Rev.}
  {\bfseries D65} (2002) 124013},
\href{http://arxiv.org/abs/gr-qc/0110035}{{\ttfamily arXiv:gr-qc/0110035
  [gr-qc]}}.

\bibitem{Maro}
D.~Marolf, ``{Quantum observables and recollapsing dynamics},''
  \href{http://dx.doi.org/10.1088/0264-9381/12/5/011}{{\em Class.Quant.Grav.}
  {\bfseries 12} (1995) 1199--1220},
\href{http://arxiv.org/abs/gr-qc/9404053}{{\ttfamily arXiv:gr-qc/9404053
  [gr-qc]}}.

\bibitem{Ditt}
B.~Dittrich, ``{Partial and complete observables for canonical general
  relativity},'' \href{http://dx.doi.org/10.1088/0264-9381/23/22/006}{{\em
  Class.Quant.Grav.} {\bfseries 23} (2006) 6155--6184},
\href{http://arxiv.org/abs/gr-qc/0507106}{{\ttfamily arXiv:gr-qc/0507106
  [gr-qc]}}.

\bibitem{PoSa}
J.~Pons and D.~Salisbury, ``{The Issue of time in generally covariant theories
  and the Komar-Bergmann approach to observables in general relativity},''
  \href{http://dx.doi.org/10.1103/PhysRevD.71.124012}{{\em Phys.Rev.}
  {\bfseries D71} (2005) 124012},
\href{http://arxiv.org/abs/gr-qc/0503013}{{\ttfamily arXiv:gr-qc/0503013
  [gr-qc]}}.

\bibitem{Giddings2005}
S.~B. Giddings, D.~Marolf, and J.~B. Hartle, ``{Observables in effective
  gravity},'' \href{http://dx.doi.org/10.1103/PhysRevD.74.064018}{{\em
  Phys.Rev.} {\bfseries D74} (2006) 064018},
\href{http://arxiv.org/abs/hep-th/0512200}{{\ttfamily arXiv:hep-th/0512200
  [hep-th]}}.

\bibitem{DiTa}
B.~Dittrich and J.~Tambornino, ``{A Perturbative approach to Dirac observables
  and their space-time algebra},''
  \href{http://dx.doi.org/10.1088/0264-9381/24/4/001}{{\em Class.Quant.Grav.}
  {\bfseries 24} (2007) 757--784},
\href{http://arxiv.org/abs/gr-qc/0610060}{{\ttfamily arXiv:gr-qc/0610060
  [gr-qc]}}.

\bibitem{GaGi}
M.~Gary and S.~B. Giddings, ``{Relational observables in 2-D quantum
  gravity},'' \href{http://dx.doi.org/10.1103/PhysRevD.75.104007}{{\em
  Phys.Rev.} {\bfseries D75} (2007) 104007},
\href{http://arxiv.org/abs/hep-th/0612191}{{\ttfamily arXiv:hep-th/0612191
  [hep-th]}}.

\bibitem{GiSl}
S.~B. Giddings and M.~S. Sloth, ``{Fluctuating geometries, q-observables, and
  infrared growth in inflationary spacetimes},''
  \href{http://dx.doi.org/10.1103/PhysRevD.86.083538}{{\em Phys.Rev.}
  {\bfseries D86} (2012) 083538},
\href{http://arxiv.org/abs/1109.1000}{{\ttfamily arXiv:1109.1000 [hep-th]}}.

\bibitem{Khavkine:2011kj}
I.~Khavkine, ``{Quantum astrometric observables I: time delay in classical and
  quantum gravity},'' \href{http://dx.doi.org/10.1103/PhysRevD.85.124014}{{\em
  Phys. Rev.} {\bfseries D85} (2012) 124014},
\href{http://arxiv.org/abs/1111.7127}{{\ttfamily arXiv:1111.7127 [gr-qc]}}.

\bibitem{Bonga:2013uha}
B.~Bonga and I.~Khavkine, ``{Quantum astrometric observables II: time delay in
  linearized quantum gravity},''
  \href{http://dx.doi.org/10.1103/PhysRevD.89.024039}{{\em Phys. Rev.}
  {\bfseries D89} (2014) 024039},
\href{http://arxiv.org/abs/1307.0256}{{\ttfamily arXiv:1307.0256 [gr-qc]}}.

\bibitem{Dirac1955}
P.~A. Dirac, ``{Gauge invariant formulation of quantum electrodynamics},''
\href{http://dx.doi.org/10.1139/p55-081}{{\em Can.J.Phys.} {\bfseries 33}
  (1955) 650}.

\bibitem{Heemskerk2012}
I.~Heemskerk, ``{Construction of Bulk Fields with Gauge Redundancy},''
  \href{http://dx.doi.org/10.1007/JHEP09(2012)106}{{\em JHEP} {\bfseries 1209}
  (2012) 106},
\href{http://arxiv.org/abs/1201.3666}{{\ttfamily arXiv:1201.3666 [hep-th]}}.

\bibitem{Kabat:2012hp}
D.~Kabat, G.~Lifschytz, S.~Roy, and D.~Sarkar, ``{Holographic representation of
  bulk fields with spin in AdS/CFT},''
  \href{http://dx.doi.org/10.1103/PhysRevD.86.026004,
  10.1103/PhysRevD.86.029901}{{\em Phys. Rev.} {\bfseries D86} (2012) 026004},
\href{http://arxiv.org/abs/1204.0126}{{\ttfamily arXiv:1204.0126 [hep-th]}}.

\bibitem{Kabat2013}
D.~Kabat and G.~Lifschytz, ``{Decoding the hologram: Scalar fields interacting
  with gravity},'' \href{http://dx.doi.org/10.1103/PhysRevD.89.066010}{{\em
  Phys.Rev.} {\bfseries D89} (2014) 066010},
\href{http://arxiv.org/abs/1311.3020}{{\ttfamily arXiv:1311.3020 [hep-th]}}.

\bibitem{Bodendorfer:2015aca}
N.~Bodendorfer, J.~Lewandowski, and J.~{\'S}wie{\.z}ewski, ``{General
  Relativity in the radial gauge I. Reduced phase space and canonical
  structure},''
\href{http://arxiv.org/abs/1506.09164}{{\ttfamily arXiv:1506.09164 [gr-qc]}}.

\bibitem{Duch:2014hfa}
P.~Duch, W.~Kaminski, J.~Lewandowski, and J.~Swiezewski, ``{Observables for
  General Relativity related to geometry},''
  \href{http://dx.doi.org/10.1007/JHEP05(2014)077}{{\em JHEP} {\bfseries 05}
  (2014) 077},
\href{http://arxiv.org/abs/1403.8062}{{\ttfamily arXiv:1403.8062 [gr-qc]}}.

\bibitem{Haag}
R.~Haag, {\em {Local quantum physics: Fields, particles, algebras}}.
\newblock (Texts and monographs in physics). Springer, Berlin, Germany,
1992.
\newblock

\bibitem{SGalg}
S.~B. Giddings, ``{Hilbert space structure in quantum gravity: an algebraic
  perspective},'' \href{http://dx.doi.org/10.1007/JHEP12(2015)099}{{\em JHEP}
  {\bfseries 12} (2015) 099},
\href{http://arxiv.org/abs/1503.08207}{{\ttfamily arXiv:1503.08207 [hep-th]}}.

\bibitem{GiLia}
S.~B. Giddings and M.~Lippert, ``{Precursors, black holes, and a locality
  bound},'' \href{http://dx.doi.org/10.1103/PhysRevD.65.024006}{{\em Phys.Rev.}
  {\bfseries D65} (2002) 024006},
\href{http://arxiv.org/abs/hep-th/0103231}{{\ttfamily arXiv:hep-th/0103231
  [hep-th]}}.

\bibitem{GiLib}
S.~B. Giddings and M.~Lippert, ``{The Information paradox and the locality
  bound},'' \href{http://dx.doi.org/10.1103/PhysRevD.69.124019}{{\em Phys.Rev.}
  {\bfseries D69} (2004) 124019},
\href{http://arxiv.org/abs/hep-th/0402073}{{\ttfamily arXiv:hep-th/0402073
  [hep-th]}}.

\bibitem{LQGST}
S.~B. Giddings, ``{Locality in quantum gravity and string theory},''
  \href{http://dx.doi.org/10.1103/PhysRevD.74.106006}{{\em Phys.Rev.}
  {\bfseries D74} (2006) 106006},
\href{http://arxiv.org/abs/hep-th/0604072}{{\ttfamily arXiv:hep-th/0604072
  [hep-th]}}.

\bibitem{Kabat2012}
D.~Kabat and G.~Lifschytz, ``{CFT representation of interacting bulk gauge
  fields in AdS},'' \href{http://dx.doi.org/10.1103/PhysRevD.87.086004}{{\em
  Phys.Rev.} {\bfseries D87} no.~8, (2013) 086004},
\href{http://arxiv.org/abs/1212.3788}{{\ttfamily arXiv:1212.3788 [hep-th]}}.

\bibitem{Bucholz1982}
D.~Buchholz, ``The physical state space of quantum electrodynamics,''
  \href{http://dx.doi.org/10.1007/BF02029133}{{\em Communications in
  Mathematical Physics} {\bfseries 85} no.~1, (1982) 49--71}.

\bibitem{Steinmann1983}
O.~Steinmann, ``{Perturbative {QED} in Terms of Gauge Invariant Fields},''
\href{http://dx.doi.org/10.1016/0003-4916(84)90053-8}{{\em Annals Phys.}
  {\bfseries 157} (1984) 232}.

\bibitem{Steinmann2004}
O.~Steinmann, ``{Physical fields in QED},'' {\em Prog.Math.} {\bfseries 251}
  (2007) 301--310,
\href{http://arxiv.org/abs/hep-th/0411095}{{\ttfamily arXiv:hep-th/0411095
  [hep-th]}}.

\bibitem{Shab}
S.~Shabanov, ``The proper field of charges and gauge invariant variables in
  electrodynamics.'' Dubna preprint JINR-E2-92-136 (unpublished).

\bibitem{HaJo}
P.~E. Haagensen and K.~Johnson, ``On the wave functional for two heavy color
  sources in Yang-Mills theory,''
\href{http://arxiv.org/abs/hep-th/9702204}{{\ttfamily arXiv:hep-th/9702204
  [hep-th]}}.

\bibitem{BLMdecomp}
M.~Lavelle and D.~McMullan, ``{Hadrons without strings},''
  \href{http://dx.doi.org/10.1016/S0370-2693(99)01326-X}{{\em Phys.Lett.}
  {\bfseries B471} (1999) 65--71},
\href{http://arxiv.org/abs/hep-ph/9910398}{{\ttfamily arXiv:hep-ph/9910398
  [hep-ph]}}.

\bibitem{Georgi1990}
H.~Georgi, ``{An Effective Field Theory for Heavy Quarks at Low-energies},''
\href{http://dx.doi.org/10.1016/0370-2693(90)91128-X}{{\em Phys.Lett.}
  {\bfseries B240} (1990) 447--450}.

\bibitem{Bagan1999a}
E.~Bagan, M.~Lavelle, and D.~McMullan, ``{Charges from dressed matter:
  Construction},'' \href{http://dx.doi.org/10.1006/aphy.2000.6048}{{\em Annals
  Phys.} {\bfseries 282} (2000) 471--502},
\href{http://arxiv.org/abs/hep-ph/9909257}{{\ttfamily arXiv:hep-ph/9909257
  [hep-ph]}}.

\bibitem{Bagan1999b}
E.~Bagan, M.~Lavelle, and D.~McMullan, ``{Charges from dressed matter: Physics
  and renormalization},'' \href{http://dx.doi.org/10.1006/aphy.2000.6049}{{\em
  Annals Phys.} {\bfseries 282} (2000) 503--540},
\href{http://arxiv.org/abs/hep-ph/9909262}{{\ttfamily arXiv:hep-ph/9909262
  [hep-ph]}}.

\bibitem{Almheiri2014}
A.~Almheiri, X.~Dong, and D.~Harlow, ``{Bulk Locality and Quantum Error
  Correction in AdS/CFT},''
\href{http://arxiv.org/abs/1411.7041}{{\ttfamily arXiv:1411.7041 [hep-th]}}.

\bibitem{Regge1974}
T.~Regge and C.~Teitelboim, ``Role of surface integrals in the Hamiltonian
  formulation of general relativity,''
  \href{http://dx.doi.org/http://dx.doi.org/10.1016/0003-4916(74)90404-7}{{\em
  Annals of Physics} {\bfseries 88} no.~1, (1974) 286 -- 318}.
  \url{http://www.sciencedirect.com/science/article/pii/0003491674904047}.

\bibitem{Ashtekar1978}
A.~Ashtekar and R.~O. Hansen, ``A unified treatment of null and spatial
  infinity in general relativity. I. Universal structure, asymptotic
  symmetries, and conserved quantities at spatial infinity,''
  \href{http://dx.doi.org/http://dx.doi.org/10.1063/1.523863}{{\em Journal of
  Mathematical Physics} {\bfseries 19} no.~7, (1978) 1542--1566}.

\bibitem{Ashtekar1992}
A.~Ashtekar and J.~D. Romano, ``Spatial infinity as a boundary of spacetime,''
  \href{http://dx.doi.org/10.1088/0264-9381/9/4/019}{{\em Classical and Quantum
  Gravity} {\bfseries 9} no.~4, (1992) 1069}.
  \url{http://stacks.iop.org/0264-9381/9/i=4/a=019}.

\bibitem{Hanson1976}
A.~Hanson, T.~Regge, and C.~Teitelboim, {\em Constrained Hamiltonian systems}.
\newblock Contributi C. linceo inter. sc. mat. Accademia Naz. dei Lincei, 1976.
\newblock \url{https://books.google.com/books?id=-u1JNAAACAAJ}.

\bibitem{FaKu}
P.~Kulish and L.~Faddeev, ``{Asymptotic conditions and infrared divergences in
  quantum electrodynamics},''
\href{http://dx.doi.org/10.1007/BF01066485}{{\em Theor.Math.Phys.} {\bfseries
  4} (1970) 745}.

\bibitem{HLM}
R.~Horan, M.~Lavelle, and D.~McMullan, ``{Asymptotic dynamics in quantum field
  theory},'' \href{http://dx.doi.org/10.1063/1.533352}{{\em J.Math.Phys.}
  {\bfseries 41} (2000) 4437--4451},
\href{http://arxiv.org/abs/hep-th/9909044}{{\ttfamily arXiv:hep-th/9909044
  [hep-th]}}.

\bibitem{Akhoury}
J.~Ware, R.~Saotome, and R.~Akhoury, ``{Construction of an asymptotic S matrix
  for perturbative quantum gravity},''
  \href{http://dx.doi.org/10.1007/JHEP10(2013)159}{{\em JHEP} {\bfseries 1310}
  (2013) 159},
\href{http://arxiv.org/abs/1308.6285}{{\ttfamily arXiv:1308.6285 [hep-th]}}.

\bibitem{SG2006}
S.~B. Giddings, ``{(Non)perturbative gravity, nonlocality, and nice slices},''
  \href{http://dx.doi.org/10.1103/PhysRevD.74.106009}{{\em Phys.Rev.}
  {\bfseries D74} (2006) 106009},
\href{http://arxiv.org/abs/hep-th/0606146}{{\ttfamily arXiv:hep-th/0606146
  [hep-th]}}.

\bibitem{GiSl2}
S.~B. Giddings and M.~S. Sloth, ``{Cosmological observables, IR growth of
  fluctuations, and scale-dependent anisotropies},''
  \href{http://dx.doi.org/10.1103/PhysRevD.84.063528}{{\em Phys.Rev.}
  {\bfseries D84} (2011) 063528},
\href{http://arxiv.org/abs/1104.0002}{{\ttfamily arXiv:1104.0002 [hep-th]}}.

\end{thebibliography}\endgroup

\end{document}